\documentclass[
superscriptaddress,
amsmath,amssymb,
aps,
prb,
twocolumn,
]{revtex4-2}

\usepackage{graphicx}
\usepackage{dcolumn}
\usepackage{bm}
\usepackage{bbold} 
\usepackage[breaklinks,colorlinks = true,linkcolor = magenta,urlcolor=magenta,citecolor=red]{hyperref}
\usepackage{url}
\usepackage{ulem}
\usepackage{tikz}
\usepackage{diagbox}
\usepackage{multirow}
\usepackage{hhline}
\usepackage{array}
\usepackage{booktabs}
\usepackage{ctable}
\usepackage{upgreek}
\usepackage{epsfig,psfrag,subfigure,amsopn}
\usepackage{mathrsfs}
\usepackage{amssymb}
\usepackage{amsbsy}
\usepackage{color}
\usepackage{cancel}
\usepackage{pifont}
\usepackage{marginnote}
\usepackage{float}
\newcommand{\bcen}{\begin{center}}
\newcommand{\ecen}{\end{center}}
\newcommand{\btab}{\begin{tabular}}
\newcommand{\etab}{\end{tabular}}
\newcommand{\bdes}{\begin{description}}
\newcommand{\edes}{\end{description}}

\newcommand{\beq}{\begin{equation}}
\newcommand{\eeq}{\end{equation}}
\newcommand{\bea}{\begin{eqnarray}}
\newcommand{\eea}{\end{eqnarray}}

\newcommand{\bary}{\begin{array}}
\newcommand{\eary}{\end{array}}
\newcommand{\benum}{\begin{enumerate}}
\newcommand{\eenum}{\end{enumerate}}
\newcommand{\bitem}{\begin{itemize}}
\newcommand{\eitem}{\end{itemize}}

\newcommand{\bS} { \mbox{\boldmath $S$}}

\newcommand{\eqn}[1] {eqn.~(\ref{#1})}

\newcommand{\Fig}[1]{Fig.~\ref{#1}}

\makeatletter

\newcommand{\Rmnum}[1]{\expandafter\@slowromancap\romannumeral #1@}
\makeatother

\begin{document}
\relax

\title{Correlation-driven non-trivial phases in single bi-layer Kagome intermetallics}

\author{Aabhaas Vineet Mallik}
\email{aabhaas@icts.res.in}
\affiliation{Department of Physics, Bar-Ilan University, Ramat Gan, Israel 5290002}
    
\author{Adhip Agarwala}
\email{adhip@iitk.ac.in}
\affiliation{Indian Institute of Technology Kanpur, Kalyanpur, Kanpur-208016}

\author{Tanusri Saha-Dasgupta}
\email{t.sahadasgupta@gmail.com}
\affiliation{Department of Condensed Matter Physics and Materials Science, S.~N. Bose National Centre for Basic Sciences, Kolkata 700098, India}

\date{\today}

\begin{abstract}
Bi-layer Kagome compounds provide an exciting playground where the interplay of topology and strong correlations can give rise to exotic phases of matter. Motivated by recent first principles calculation on such
systems (Phys. Rev. Lett {\bf 125}, 026401), reporting stabilization of a Chern metal with topological nearly-flat band close to Fermi level, we build minimal models to study the effect of strong electron-electron interactions on such
a Chern metal. Using approriate numerical and analytical techniques, we show that the topologically non-trivial bands present in this system at the Fermi energy can realize fractional Chern insulator states. We further show that if the time-reversal symmetry is restored due to destruction of magnetism by low dimensionality and fluctuation, the system can realize a superconducting phase in the presence of strong local repulsive interactions. Furthermore, we identify an interesting phase transition from the superconducting phase to a correlated metal by tuning nearest-neighbor repulsion. Our study uncovers a rich set of non-trivial phases realizable in this system, and contextualizes the physically meaningful regimes where such phases can be further explored. 
\end{abstract}

\maketitle
\clearpage
\newpage

\section{Introduction}

The Kagome lattice -- built out of corner sharing triangles -- presents a rather interesting situation where both the itinerancy of the electrons as well as the effect of electron-electron interactions, can be frustrated. The frustration of the itinerant electrons is manifested through the characteristic flat / nearly-flat bands in the various short range tight-binding models on the Kagome lattice. When the Fermi energy lies in one of these bands of narrow band-width then the electron-electron interactions are expected to play a crucial role in determining the ground state properties of the system. Together with this effect,  the presence of spin-orbit coupling in the nearly-flat non-interacting band at the Fermi energy may lead to a non-trivial band topology~\cite{Hasan_RMP_2010, Chiu_RMP_2016, Ludwig_PS_2015, Qi_RMP_2011, Haldane_PRL_1988, KaneMele_PRL_2005}. This interplay of electron-electron interactions and band topology poses outstanding challenges and has been of much interest, for example, in the context of fractionally filled Landau levels in the quantum Hall systems \cite{Jain2007composite, HanssonRMP2017,Murthy_RMP_2003} and Moire graphene \cite{Bistritzer_PNAS_2011, Cao_Nature_2018} more recently. Interestingly, a plethora of recently discovered metallic systems based on the Kagome motif \cite{ Caer_1978,Nakatsuji_Nature_2015,Ye_Nature_2018, Kiyohara_PRA_2016, nayak2016large, Yan_APL_2019, kuroda2017evidence, Zhiyong_PRL_2018, yin2019negative, kang2020dirac, Chen_FP_2021, ZENG2021765,Sales_PRM_2021, WangPRB2021, YuPRB2021, ZhangPRB2021, Shi_arXiv_2021, 2021arXiv210710714Y, 2021arXiv210613443M, Li_PRL_2019, Ye_arXiv_2021} provide a wide material basis to realize and explore this physics further.

\begin{figure}
\includegraphics[width=0.8\columnwidth]{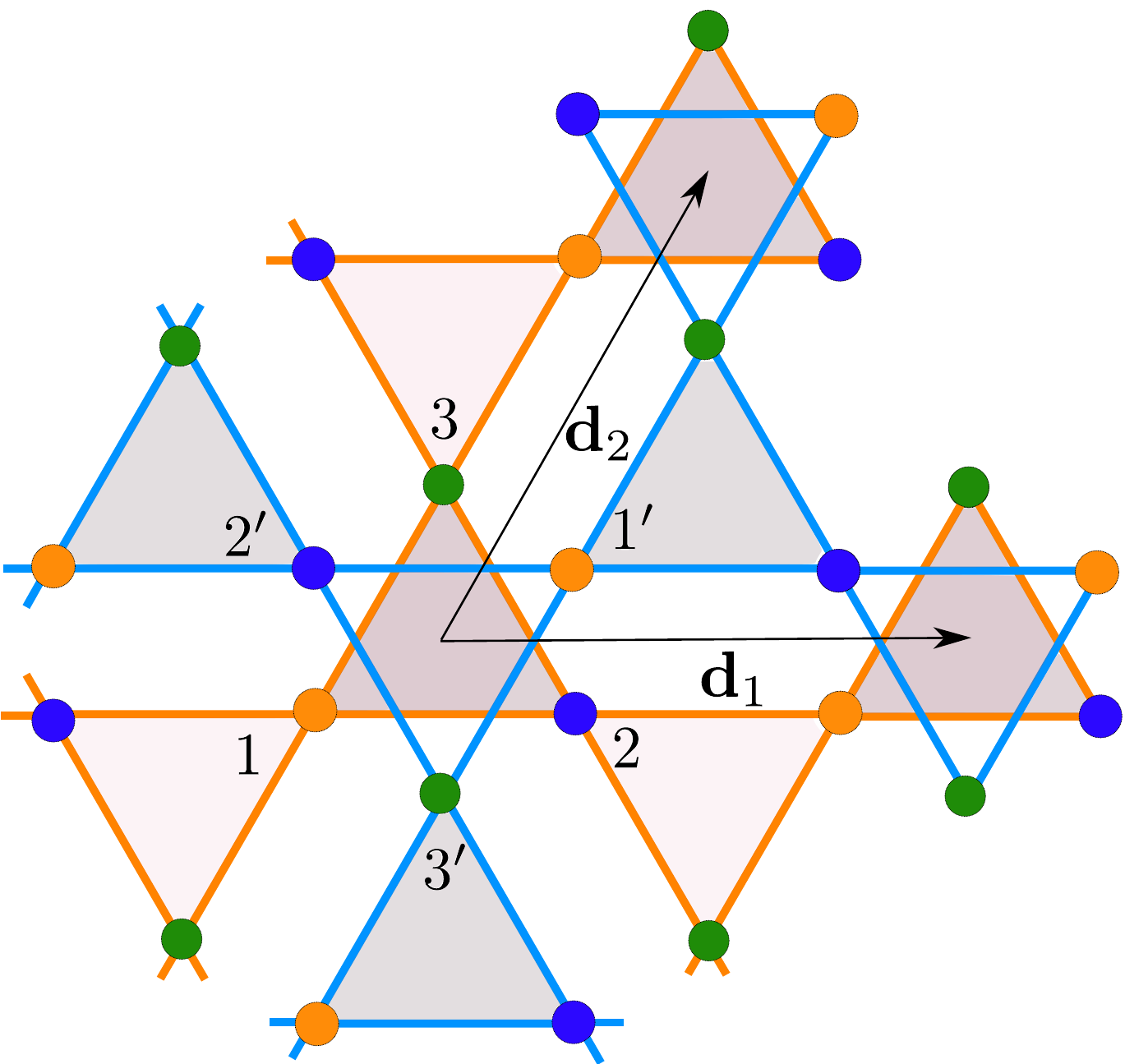}

\caption{{\it Bi-layer Kagome lattice:} A Kagome bi-layer lattice has a tripartite structure with a unit-cell containing six identical atoms labelled (1-3) (bottom-layer) and ($1'$-$3'$) (top-layer) and shown by different colors. Inter and intra-unit cell nearest-neighbor hoppings can have different hopping strengths due to breathing anisotropy.  The lattice vectors are given by ${\bf d_1} = \left\{ 1,0 \right\},~~~{\bf d_2} = \left\{ \frac{1}{2},\frac{\sqrt{3}}{2} \right\}$. The reciprocal lattice vectors are given by ${\bf b_1} = \left\{\frac{\sqrt{3}}{2}, -\frac{1}{2} \right\}\frac{4\pi}{\sqrt{3}},~~
{\bf b_2} = \left\{ 0,1 \right\}\frac{4\pi}{\sqrt{3}}$.}
\label{fig:figure1} 
\end{figure}

A particularly interesting family of Kagome based metallic systems occur in the binary intermetallics $M_3$Sn$_2$, where $M$ = Mn, Fe, Ni, Cu, Co represents a transition metal\cite{Caer_1978,Nakatsuji_Nature_2015,Ye_Nature_2018, Kiyohara_PRA_2016, nayak2016large, Yan_APL_2019, kuroda2017evidence, Zhiyong_PRL_2018, yin2019negative, kang2020dirac, Chen_FP_2021, ZENG2021765, lyalin2021spin, neupert2021charge,YangPRB2022}. These materials form a three dimensional layered structure with the basic motif being a bi-layer Kagome of the $M$ atoms, as shown in Fig. 1. The sought-after flat bands have been observed in iron compounds \cite{kang2020dirac}, in cobalt systems \cite{liu2020orbital,kang2020topological}, and most recently in manganese based intermetallics \cite{li2021dirac, yin2020quantum}. The interplay of strong correlations and topological properties is manifested in a wide range of observed exotic phenomena, such as the co-existence of magnetism and anomalous Hall effect seen recently in manganese systems (along with rare earths) \cite{Lingling_APL_2021}, an exotic charge density wave order and superconductivity seen in related antimony compounds \cite{WangPRB2021, YuPRB2021, ZhangPRB2021, yu2021unusual, LiPRX2021, 2023arXivTuniz}. A variety of other observations including skyrmions and topological Hall effect \cite{DuPRL2022}, hole pockets at the Fermi surface \cite{2022arXivEkahana}, and spin waves \cite{2023arXivZhang} have further made the field extremely intriguing.

An especially intriguing system in this family are the FeSn compounds where the stacking arrangement can alter the nature of both the magnetic properties of the ground state as well as that of the itinerant electrons. For instance, when the two Kagome Fe$_3$Sn layers in a bi-layer motif are aligned such that the Fe atoms in the two layers are right on top of each other, the system is antiferromagnetic with strong spin fluctuations and localized electrons \cite{Zhang_PRB_2022, multer2023imaging, li2022spin}. On the other hand when the two Fe$_3$Sn layers and aligned such that Fe atoms in the two layers form a star of David (see \Fig{fig:figure1}), the system is ferromagnetic with a flat band near the Fermi energy which is topological in nature making it susceptible towards a host of exotic phenomena \cite{Ye_Nature_2018, khan2022intrinsic, Baidya_PRL_2020}. Given the susceptibility of this latter system to a host of exotic phases it forms the focus of our study in this paper. More specifically, we focus on the single bi-layer limit. As shown in one of our previous work\cite{Baidya_PRL_2020} cleavibility of these compounds, allows
possible synthesis of bi-layer and its tunability, for example, by using different substrates \cite{lyalin2021spin,khan2022intrinsic}. 
The situation is further promising as such epitaxial films have recently been synthesized \cite{ChengarXiv2021}.

Although, the experiments and first principles calculation for the bulk Fe$_3$Sn$_2$ \cite{Ye_Nature_2018, Baidya_PRL_2020} reveal a complicated band structure close to the Fermi energy, interestingly, in the single bi-layer limit, a simplified low energy band structure separated from the high energy bands emerges. This allows for the possibility of analysing the low energy physics of the single bi-layer systems using effective tight-binding models with short-range electron-electron interactions~\cite{Baidya_PRL_2020}. 
In the bi-layer limit, the first-principles calculation\cite{Baidya_PRL_2020} found the Fe based system to be a Chern metal,
while the Ni/Co based system to be nonmagnetic. With these first-principles inputs, it will be a worth-while exercise to understand the possible interaction-driven instabilities towards non-trivial phases in the nearly-flat bands of the spin-polarized Chern metal as well as the nonmagnetic metal.

\begin{figure}
    \centering
\includegraphics[width=0.8\columnwidth]{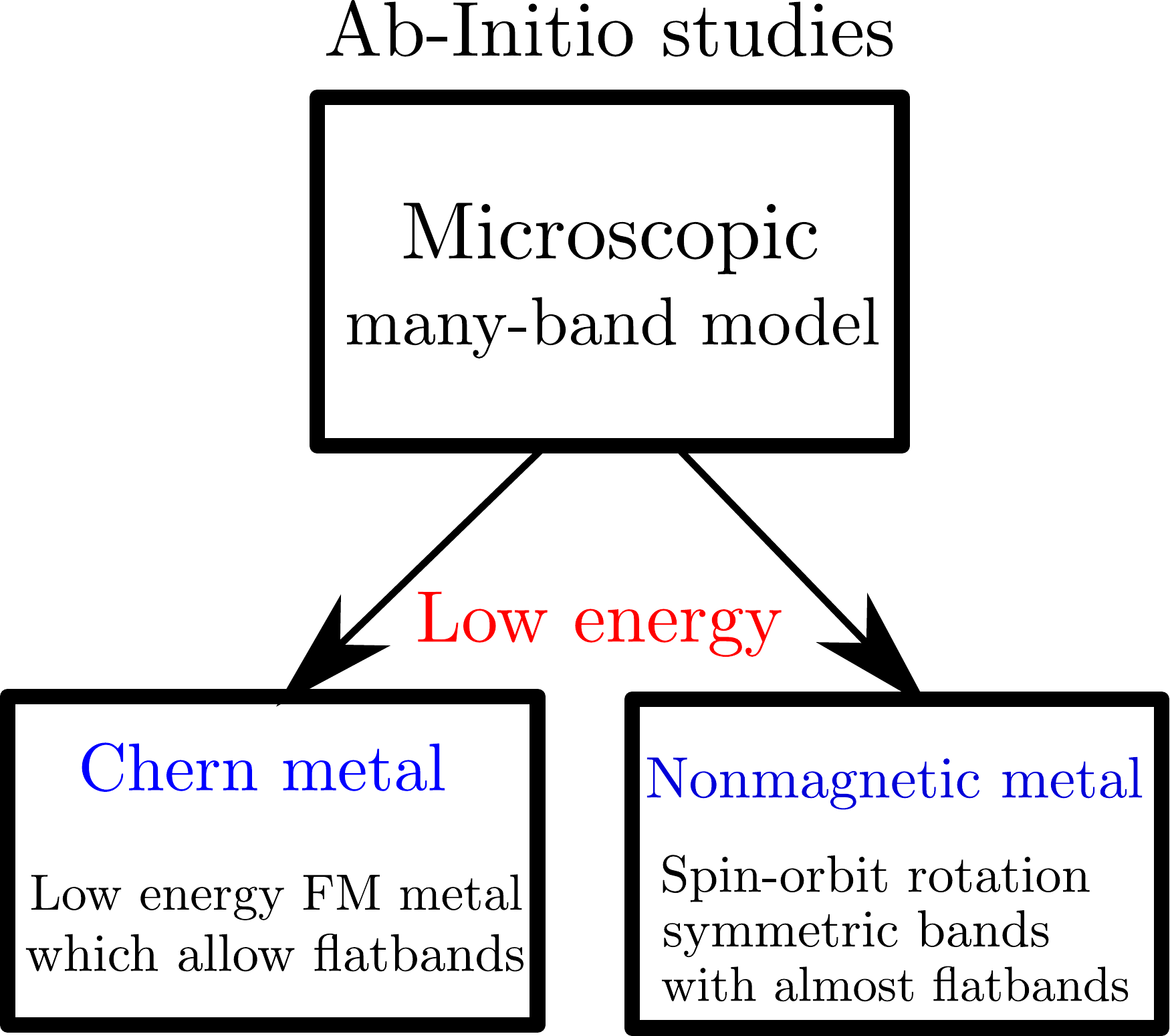}
    \caption{{\it Effective low energy physics:} Ab-initio studies lead to two broad category of microscopically interesting systems (i) Chern metals -- which spontaneously breaks time-reversal symmetry and have a partially filled flatband with a non-zero Chern number, (ii) Nonmagnetic metals -- a metallic phase which retains the complete spin-orbit rotation symmetry.}
    \label{fig:flowchart}
\end{figure}

In this paper, we take up the above task within the framework of
a minimal symmetry allowed hopping Hamiltonian, characterized by inter-layer hybridization, intra-layer breathing distortion due to differently sized up and
down triangles, and inter-layer potential difference. The microscopic many-body
model built up on this, hosting flat-bands is studied for both a) a low-energy Chern metal and b) a low-energy nonmagnetic metal, as shown in Fig.~\ref{fig:flowchart}. Our analysis establishes that possible tuning of these parameters by substrate effect, strain as well as gating, can give rise to a plethora of correlation-driven exotic phases. For the Chern metal, when the two Kagome layers are weakly hybridized, the repulsive interactions in the ferromagnetic flat-band can stabilize a 1/3 fractional Chern insulator (FCI) state, while possibly stabilizing a 1/5 FCI state in the limit of strong interlayer hybridization. On the other hand, in the case of
non-magnetic metal, we discover a pairing instability driven by onsite repulsion within a $t$-$J$ model calculation. The transition temperature of the corresponding superconducting
phase can be as high as about 30$K$. Upon varying the strength of the repulsive
interactions, a transition to a correlated metallic phase, with the promise of
hosting spin liquid phase has been discovered.

While our work is motivated by the bi-layer physics of
Fe$_3$Sn$_2$, the analysis presented here is general in nature and should be applicable to a range of bi-layer Kagome thin films.

\section{Low energy Hamiltonians for bi-layer Kagome metals}
\label{sec:FeSn}

A minimal symmetry allowed hopping Hamiltonian, consisting of one Kramers doublet per site (\Fig{fig:figure1}), is given by
\begin{align}
    H_{tb}=\sum_{ij,\sigma}t_{ij}^{\sigma} c_{i\sigma}^\dagger c_{j\sigma}-\mu\sum_{i} n_i
    \label{oneorbHam}
\end{align}
where, $i$ and $j$ refer to the sites on the bi-layer Kagome lattice, $c_{i\sigma}$ ($c_{i\sigma}^\dagger$) are the electron annihilation (creation) operators at site $i$ with spin $\sigma=\uparrow,\downarrow$ and $n_i=\sum_\sigma c^\dagger_{i\sigma}c_{i\sigma}$ is the onsite electron number operator. $t _{ij}^{\sigma}$ denote various hopping amplitudes, including an effective spin-orbit coupling term, and $\mu$ is the chemical potential.

In this paper, we shall use this single orbital hopping Hamiltonian $H_{tb}$ to study the competition between various relevant phases. Such an approach has proved to be fruitful in understanding a host of flatband systems, e.g.~in various Moire structures \cite{Giem_NatMat_2008}. It is worth noting that the above minimal model has several parameters which in principle can be tuned in the experiments on the bi-layer Kagome materials. Below we list the important model parameters and mention the range of their expected tunablility from studies on closely related material systems.

\begin{itemize}

    \item  The nearest neighbor hopping amplitude $t$ (~$\sim0.1eV$ for Fe$_3$Sn$_2$\cite{Baidya_PRL_2020}, $\sim 0.2 eV$ for CoSn \cite{kang2020topological}) can possibly be tunned using strain \cite{Yin_PRB_2021}.

    \item The chemical potential $\mu$ which controls the filling and can be changed, for example, using gating techniques and chemical doping \cite{Sales_PRM_2021}.
    
    \item A relative potential difference between the two layers $D$ breaks the inversion symmetry of the system and can be used to model the choice of the substrate. Such effects may be important in epitaxially grown films \cite{fujiwara2021tuning}.
    
    \item The symmetry consistent breathing anisotropy between the up and the down triangles of each Kagome layer, shown as two differently shaded triangles in Fig. 1, which gives rise to two different sized triangles. This effect is incorporated by scaling the intra-layer inter-unit cell hopping amplitude by a factor $r$ compared to the intra-unit cell hopping in the same layer.
    In experiments the ratio of the intra-unit cell and inter-unit cell bond lengths can be $\sim 0.7$ in some Kagome compounds \cite{Li_PRB_2021} and can also be tuned by applying anisotropic pressure \cite{Consiglio_arXiv_2021} or by the use of appropriate substrates \cite{Zhang_PRB_2021}.
    
    \item The hopping amplitudes $t_\perp$, $t_{\perp 1}$ and $t_{\perp 1}'$ (details in Appendix~\ref{appen_tb}) model the various short-ranged inter-layer hybridization strengths. While, $t_\perp$ is expected to be around ($\sim 0.3t$) in bi-layer FeSn compounds \cite{Crasto_PRB_2019,Ye_Nature_2018}, it can be tuned by the application of pressure \cite{Consiglio_arXiv_2021} or by applying an uniaxial strain.
    
    \item We model the effect of atomic spin-orbit coupling, expected to be present in these intermetallics \cite{Baidya_PRL_2020}, by including a nearest neighbor intra-layer hopping process with an imaginary amplitude
    \begin{align}
        i \lambda (c_{i\uparrow}^\dagger c_{j\uparrow} - c_{i\downarrow}^\dagger c_{j\downarrow}),
        \label{eq_soc}
    \end{align}
    where $j\rightarrow i$ has an anti-clockwise orientation over any triangular motif. In a related system such a $\lambda$ is known to be $\sim 30-40 meV$ \cite{Crasto_PRB_2019,Ye_Nature_2018, Fang_PRB_2022} and about $\sim 0.1$eV in CoSn \cite{kang2020topological}.
    
\end{itemize}

\begin{figure}
\centering
    \includegraphics[width=0.95\columnwidth]{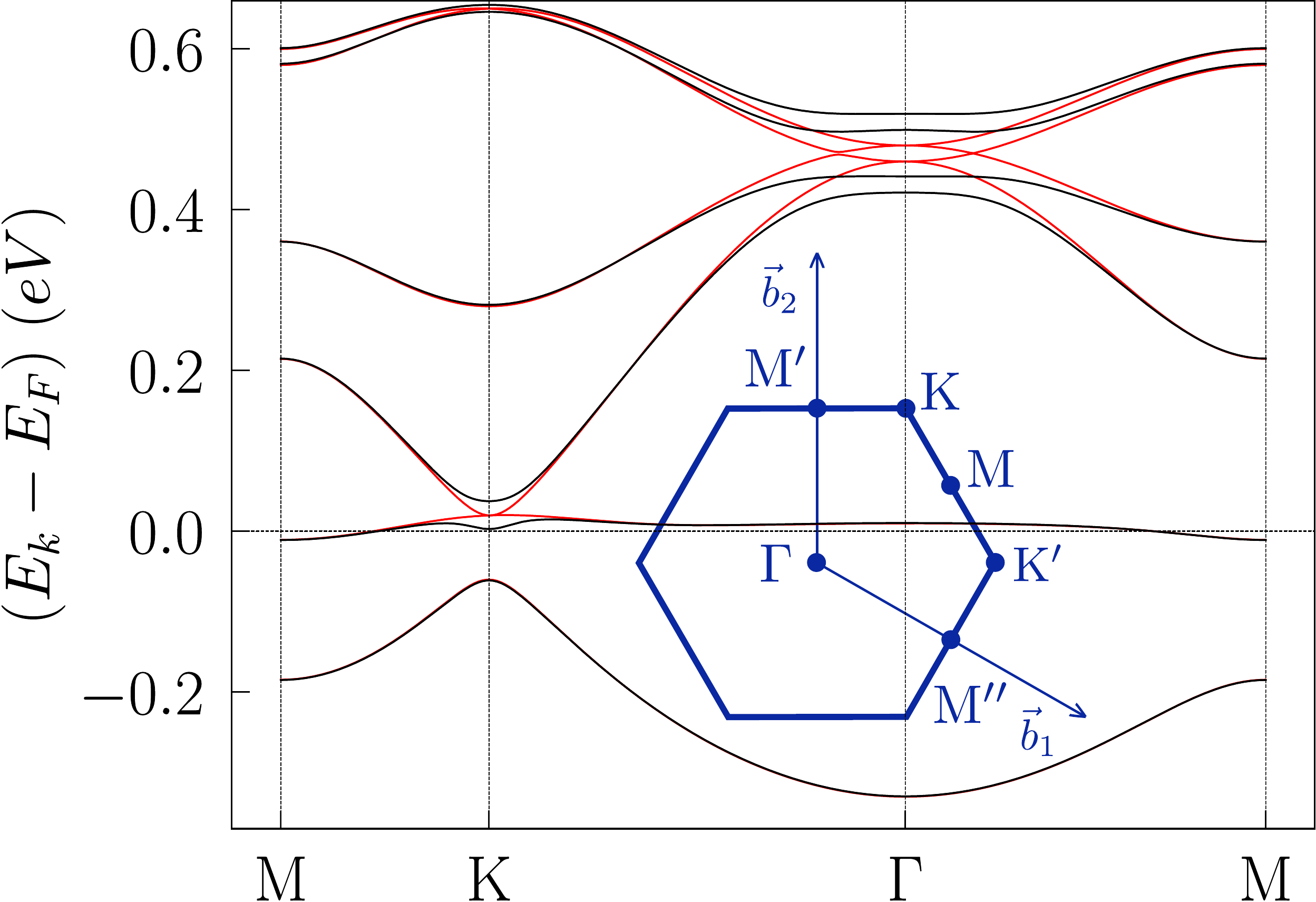}
\caption{{\it Representative band structure:}	A representative band structure of the
 minimal symmetry allowed hopping Hamiltonian, with the choice of $t=0.12$ eV, $r=1.25$, $t_{\perp}=0.06$ eV, $t_1=-0.03$ eV, $t_{\perp 1}=0.05$ eV, $\mu=-0.26$ eV and $\lambda=0.01$ eV (see text for further details). 
 The corresponding band structure switching off the spin-orbit coupling ($\lambda=0$) is shown for comparison (in red).
 The inset shows the high symmetry points and the reciprocal lattice vectors in a hexagonal Brillouin zone.}
	\label{fig:lambda_bands}
\end{figure}

Further details of the hopping Hamiltonian and its various parameters are provided in Appendix~\ref{appen_tb}. Before moving on to discuss the effect of electron-electron interactions, we would like to discuss the salient features of such a symmetry consistent tight binding Hamiltonian. A representative band structure obtained within our one-orbital-per-site model is shown in Fig.~\ref{fig:lambda_bands}. Most remarkable is the occurrence of an isolated nearly {\it flat band} at the Fermi energy, which captures the large density of states also seen from the first principles results for Fe$_3$Sn$_2$ \cite{Baidya_PRL_2020}. In the absence of spin-orbit coupling $\lambda$ (Eq.~\ref{eq_soc}) the nearly-flat band touches the dispersing band at the $K$ point of the Brillouin zone (BZ). This degeneracy, however, is generically lifted for any finite $\lambda$ owing to the difference in the symmetries of the model Hamiltonian in the presence and in the absence of the $\lambda$ term (details in Appendix \ref{appen_tb}). In other words, for a non-zero $\lambda$ our minimal tight binding model is generically expected to host an isolated nearly-flat band. In the next section we discuss the role of interactions on this nearly-flat tight-binding band.

{\subsection{Interactions}
\label{sec:ferromagnetism}

Short-ranged microscopic interactions on the low energy electrons are given by a generalised Hubbard model of the form

\begin{align}
    H_{\rm int}=U \sum_{i} n_{i \uparrow}n_{i \downarrow}+\sum_{ij}V_{ij}n_{i} n_{j}
    \label{eq:H_int}
\end{align}
where $n_{i\uparrow(\downarrow)}$ is the electron number operator at site $i$ with spin up (down), while $n_{i} = n_{i\uparrow} + n_{i\downarrow}$ is the total electron number operator at the site $i$. $U$ is the onsite repulsive interaction strength and $V_{ij}$ represent the off-site nearest neighbour (both intra and inter-layer) density-density interactions between the electrons at sites $i$ and $j$.

\subsubsection*{The Ferromagnetic Chern metal}

First principles calculations suggest that the bi-layer Fe$_3$Sn$_2$ should stabilize a ferromagnetic ground state~\cite{Baidya_PRL_2020}. Such a ferromagnetic state can indeed be favored by the onsite term in \eqn{eq:H_int} as is demonstrated within the following mean-field analysis \cite{fradkin2013field},

\beq
 U n_{i\uparrow }n_{i \downarrow} = \frac{U}{2} \Big( n_{i  \uparrow} + n_{i \downarrow} - \frac{4}{3} (\bS_{i}\cdot \bS_{i })  \Big)\rightarrow -\frac{4U}{3} {\bf m}\cdot {\bf S}_i+\cdots
 \label{eq:U_m}
\eeq
where $S_{i}^{(\tau)} = \frac{1}{2} \sum_{ss'} c^\dagger_{i s}\sigma^{(\tau)}_{s s'} c_{i s'}$ are the electron spin operators with $\sigma^{(\tau)}$ being the the Pauli matrices and ${\bf m}=\langle {\bf S}_i\rangle$ is the uniform mean-field magnetisation which we take to be along the $z-$axis, {\it i.e.} $m_z  = \frac{1}{N}\sum_{i} \langle S^z_{i} \rangle \neq 0 $, $ m_x = m_y  =  0 $. While ignoring the effect of the sub-dominant nearest neighbor interactions $V_{ij}$, self-consistent mean-field calculations for tight binding parameters presented in Fig.~\ref{fig:lambda_bands} indicate that $U\gtrsim t$ is sufficient to completely spin polarize the bands close to the Fermi energy (see Appendix~\ref{sec:mft_fm}).

A few representative band structures deep in the ferromagnetic phase, where the bands close to the Fermi energy are fully spin polarized, are presented in \Fig{fig:1byn_bands}. \Fig{fig:1byn_bands}(a) shows a representative example in the weak inter-layer hybridization limit (see caption for the parameters), while \Fig{fig:1byn_bands}(b) shows an example in the strong inter-layer hybridization limit (see caption for the parameters).  Remarkably, in both these limits, the resultant low energy bands are endowed with non-trivial Chern numbers ($C$) as shown in the figure. These two classes of band structures are robust over a finite window of $\lambda\neq 0$ as well as other parameters. The two $C=1$ bands in the weak coupling limit, can almost exclusively be associated with one of the Kagome layers \cite{Trescher_PRB_2012}; one band from each of the Kagome layers. Interestingly, however, on increasing the inter-layer hybridisation, the two $C=1$ bands undergo a Dirac gap closing at the $K'$ point (see \Fig{fig:1byn_bands}(c)). With further increase in the inter-layer hybridisation the gap reopens giving rise to a topologically trivial band and a $C=2$ band at the Fermi level.

In what follows we shall devote our attention to the above weak and strongly hybridized low energy Chern bands of reduced bandwidth to explore the effect of interactions on the Chern metal. We note, however, that strong quantum fluctuations arising specially from reduced dimensionality of the bi-layer may destabilize the ferromagnetic metal in favor of a nonmagnetic one. This provides a completely different starting point for exploring the effect of electron-electron interactions on unpolarised non-topological flat bands. In the next couple of sections we shall consider these two different classes separately and investigate the interesting phases that can possibly be realized.

\section{Fractional Chern Insulating Phases}
\label{sec:FCI}

For the electrons in the spin-polarised nearly-flat Chern band of the ferromagnetic metal, short range interactions can lead to a host of novel phases some of which we explore in this section. The electronic Chern metal has an instability for attractive nearest neighbor interactions which stabilizes a topological superconductor, as was previously discussed in \cite{Baidya_PRL_2020}.  We now discuss the instabilities of the Chern metal to repulsive interactions such as the nearest neighbor density-density repulsion in \eqn{eq:H_int}. A simple analysis suggests that such interactions cannot stabilize a superconducting phase within the mean-field approximation (see Appendix~\ref{SCspinpola} for a discussion). In this section we focus our investigation towards the possibility of realizing the fractional Chern insulator (FCI) states driven by these nearest neighbor density-density interactions in the Kagome bi-layer.

\begin{figure}
\centering
\includegraphics[width=0.99\columnwidth]{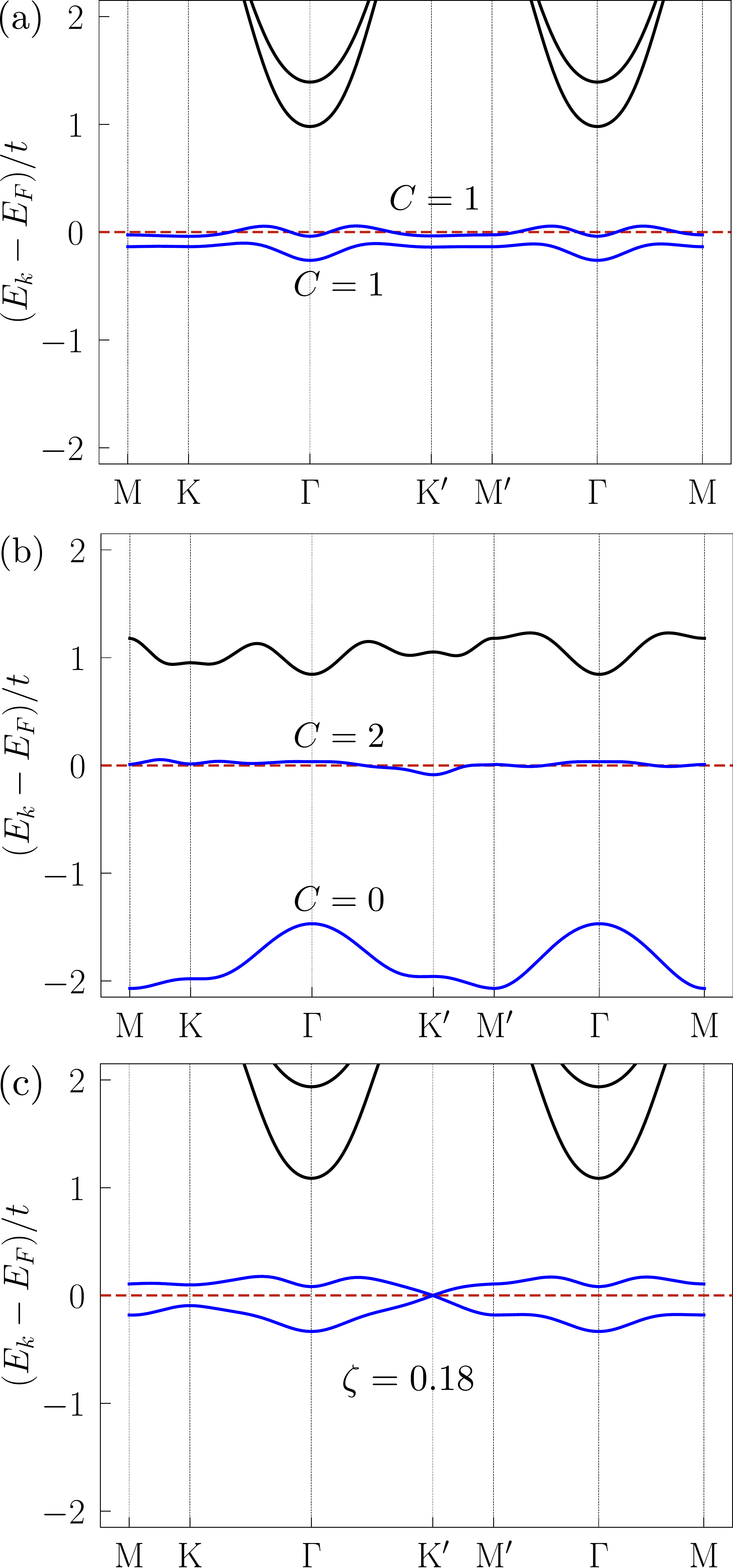}
\caption{{\it Weak and strong hybridization:} (a) The low energy tight binding bands in the weak inter-layer coupling limit ($r=1.0$, $t_{\perp}/t=0.1$, $t_{1}/t=-0.2$, $t_{\perp 1}/t=-0.1$, $t'_{\perp 1}/t=0.0$, $\lambda/t=1.0$ and $D/t=0.1$), denoted by Hamiltonian, $H_I$. (b) The low energy tight binding bands in the strong inter-layer coupling limit ($r=0.9$, $t_{\perp}/t=0.7$, $t_{1}/t=-0.3$, $t_{\perp 1}/t=0.25$, $t'_{\perp 1}/t=-0.5$, $\lambda/t=0.3$ and $D/t=0.1$), denoted by Hamiltonian, $H_{II}$. (c) The Low energy bands for $H_\zeta \equiv H_{I} + \zeta (H_{II}-H_{I})$ ($\zeta \sim 0.18$), exhibiting a Dirac crossing at the $K'$ point.}
\label{fig:1byn_bands}
\end{figure}

Robust FCI states often arise as an interaction driven instability of a fractionally filled Chern band at a particular filling of $1/(2m+1)$ with $m$ being an integer~\cite{Regnault_PRX_2011}. Moreover, for a topological band with Chern number $C$ one expects the FCI states at an electronic filling of $1/(2C+1)$ to be the most robust \cite{1991_PRB_Fradkin}. This expectation is borne out by several numerical studies on different lattice systems \cite{Wang_PRB_2012, Bergholtz_IJMPB_2013, Lauchli_PRL_2013}, and as we show below, it holds good for our bi-layer Kagome as well.

To examine this instability we perform exact diagonalization (ED) studies while incorporating {\it both} intra-layer nearest neighbor interactions $\equiv V_{\parallel}$ and inter-layer nearest neighbor interactions $\equiv V_{\perp}$. Proceeding conventionally \cite{Regnault_PRX_2011}, we drop the dispersion of the band at the Fermi energy and project the interactions to just this flat band while ignoring the Hartree terms from the projected interactions. These approximations allows us to distill signatures of the FCI states within our ED studies.  Our ED studies are performed in the momentum space where the Hamiltonian is block diagonal in the total momentum \cite{Regnault_PRX_2011}. We are, however, limited to a momentum grid of $\sim 35$ points for which the low energy eigen-spectrum can be obtained efficiently.

The considerations described above lead to the following effective Hamiltonian
\beq
H_{\text{int}} = {\cal P}\sum_{k_1,k_2,k_3,k_4} V^{\alpha_1,\alpha_2,\alpha_3,\alpha_4}_{k_1,k_2,k_3,k_4} c^\dagger_{k_1,\alpha_1} c^\dagger_{k_2,\alpha_2} c_{k_3,\alpha_3} c_{k_4,\alpha_4} {\cal P}\ .
\label{proj}
\eeq
where $\alpha$'s run over the six bands of our spin-polarized bi-layer Kagome system and ${\cal P}$ implements the projection to the band at the Fermi energy. For the results presented in this section we shall set $V_{\perp}=V_{\parallel}=1$ eV while noting that small deviations from this choice do not destabilize the FCI phases that we uncover.
Given the tunability of the tight binding parameters (discussed in Sec.~\ref{sec:FeSn}) we can certainly realize much flatter bands with non-trivial Chern numbers which are neatly separated from the other bands in the system. Thus, we believe that the FCI states that we uncover are representatives of similar and possibly more robust FCI states realizable in distinct tight binding parameter regimes of the bi-layer Kagome system.

\subsubsection{\texorpdfstring{$1/3$}{} Fractional Chern Insulator}

In the weak interlayer hybridization limit of Fig. \ref{fig:1byn_bands}(a) we project the interactions on to the second lowest (energy) Chern band with C=1. At 1/3rd filling of this band our exact diagonalization (ED) of $H_{\text{int}}$ (eqn.~\ref{proj}) reveals compelling characteristic features of the $1/3$ FCI state.

With periodic boundary conditions we find a threefold quasi-degenerate ground state manifold and a finite gap to excitations (see \Fig{fig:1by3state}(a)). When a flux $\Phi$ is introduced via twisted boundary conditions the threefold quasi-degenerate ground state manifold remains separated from the excited states while exhibiting a twist angle periodicity of $6\pi$, confirming their FCI character (see \Fig{fig:1by3state}(b)) \cite{Regnault_PRX_2011}. We also find that for the states in the ground state manifold $\langle n_k \rangle $, occupancy of the single particle Bloch state with crystal momentum $k$ in the band at the Fermi energy, equals $\sim 1/3$ (see \Fig{fig:1by3state}(c)). This again, is as expected for an incompressible $1/3$ FCI state \cite{Regnault_PRX_2011}.

While a $1/3$ FCI state in a single layer Kagome system has been reported earlier, for example in Ref.~\onlinecite{Regnault_PRX_2011}, our results in Fig.~\ref{fig:1by3state} show that such a FCI state is stable even in the {\it bi-layer} system in the presence of inter-layer hybridization ($t_\perp$) and appreciable inter-layer repulsion ($V_\perp$). 

\begin{figure}
    \centering
    \includegraphics[width=1.0\columnwidth]{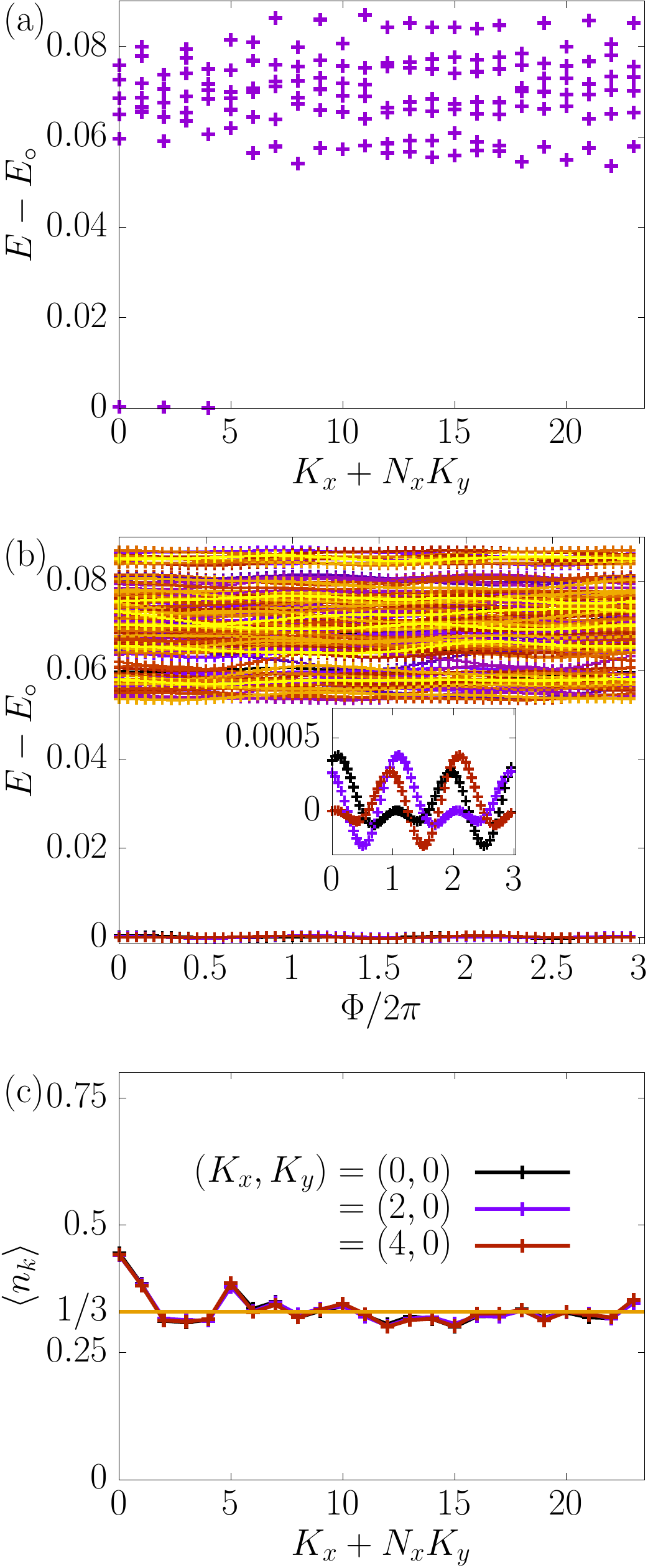}
    \caption{{\it FCI 1/3 state:} (a) Low energy spectra for the 
    $1/3$ state on a 6$\times$4 periodic boundary system for the bands in \Fig{fig:1byn_bands}(a) showing a three fold degenerate ground state.  (b)
    Flux threading results for the $1/3$ state showing that the ground state manifold remains gapped while exhibiting a periodicity of $\Phi = 6\pi$. Different colors are used to distinguish the total momenta of the states. (c) $\langle n(k) \rangle \sim 1/3$ for all values of $k$ in the three ground states.}
    \label{fig:1by3state}
\end{figure}

\subsubsection{\texorpdfstring{$1/5$}{} Fractional Chern Insulator}

We now investigate the effect of nearest neighbor repulsion on the Chern metallic phase, in the strong hybridization limit
with the Fermi energy in the $C=2$ band, shown in \Fig{fig:1byn_bands}(b). Remarkably, at $1/5$ filling of this band, within the exact diagonalization scheme discussed above, we realize a $1/5$ FCI state. With periodic boundary conditions, this FCI phase is characterized by a five fold quasi-degenerate ground state manifold (see \Fig{fig:1by5}(a)) which is clearly separated from the excited states by a gap that does not change significantly over the few system sizes that we could explore. Moreover, the number of low energy quasi-hole states expected for the $1/5$ FCI \cite{Regnault_PRX_2011}, exactly matches the number of states that our computation yields (see \Fig{fig:one_qh_1by5} of Appendix~\ref{FCIextra}).  Furthermore, with twisted boundary conditions where an additional flux $\Phi$ is introduced, one finds that the five states in the ground state manifold remain isolated from the excited states while exhibiting a twist angle periodicity of $10\pi$ (see \Fig{fig:1by5}(b)); yet another important characteristic of the $1/5$ FCI phase.

The average density of electrons in a single particle Bloch state having Bloch momentum $k$ in the band at the Fermi energy, $\langle n_k \rangle$, is shown in \Fig{fig:1by5}(c) for all the five states in the quasi-degenerate ground state manifold. While four of the states show $\langle n(k) \rangle \sim 1/5$ signaling a uniform density, 
the lowest energy state exhibits distinctive oscillations over the mean density of 1/5. We find such density oscillations in the lowest eigenstate for the two system sizes accessible within our ED studies ($5\times 6$ and $5\times 7$) which indicates a weak breaking of the rotational symmetry. While this may be an artifact of the lattice sizes and aspect ratios accessible in our ED studies, such states, if stable even in the thermodynamic limit, may also be relevant for the recently observed rotation symmetry breaking seen in a related Kagome material KV$_3$Sb$_5$\cite{2022_NatPhys_Li,2022_Nature_Nie,2020_arXiv_Jiang}. 

It is worth noting that the $1/5$ FCI state that we uncover for the bi-layer Kagome system is reminiscent of a similar state which was found for pyrochlore slabs that also inherits the Kagome motifs \cite{Trescher_PRB_2012}. We also note that for a $C=2$ tight binding band which closely resembles the band at the Fermi energy for the bi-layer Fe$_3$Sn$_2$ as obtained from the first principles, we do not find a FCI state for the system sizes that we could access.

\begin{figure}
    \centering
    \includegraphics[width=1.0\columnwidth]{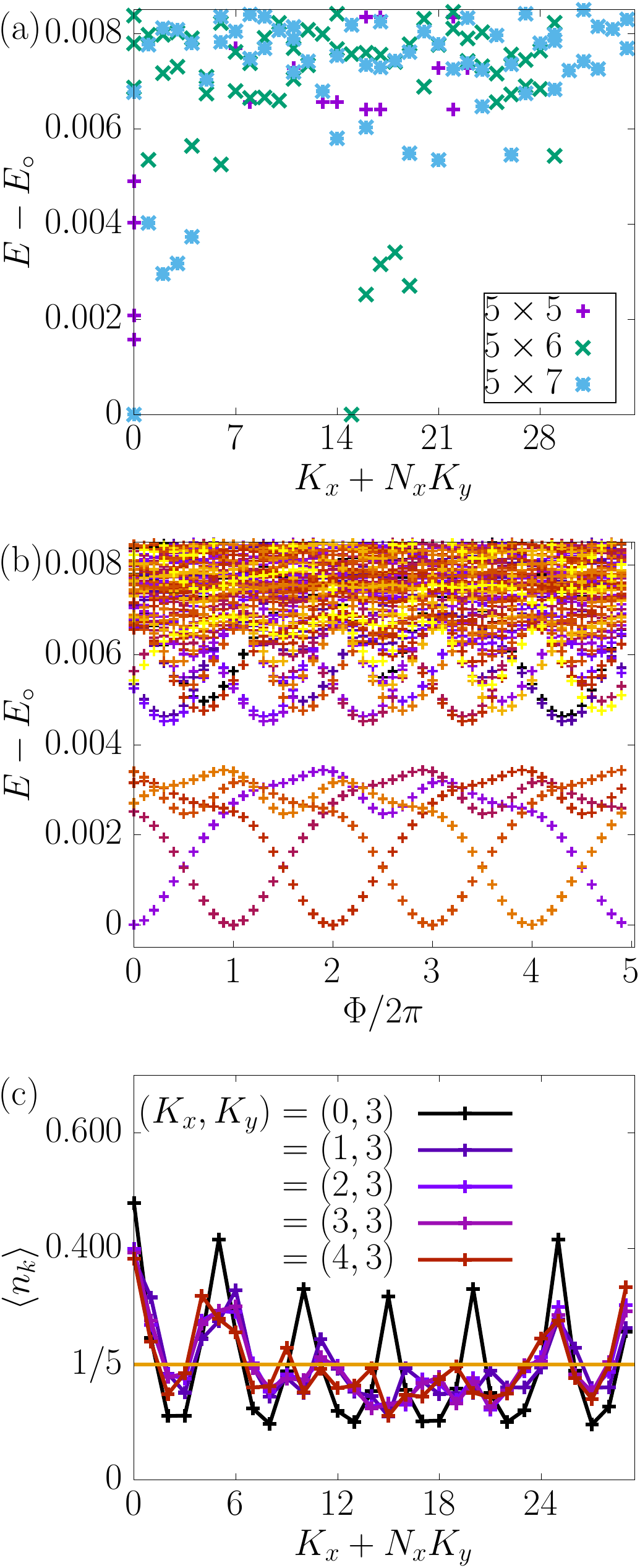}
    \caption{{\it FCI 1/5 state:} (a) Low energy spectra for the 
    $1/5$ state with periodic boundary system for the band in \Fig{fig:1byn_bands}(b) showing a five fold degenerate ground state manifold for different system sizes. (b) Flux threading results for the $1/5$ state on a 5$\times$6 momentum space grid showing that the ground state manifold remains gapped while exhibiting a periodicity of $\Phi = 10\pi$. Different colors are used to distinguish the total momenta of the states. (c) $\langle n(k) \rangle \sim 1/5$ for all $k$ in the five quasi-degenerate ground states for the 5$\times$6 system.}
    \label{fig:1by5}
\end{figure}

\section{Instabilties of the non-magnetic phase}
\label{sec:PMphase}

Having discussed the phases in the vicinity of the Chern metal, we now turn to the case where the quantum fluctuations in the bi-layer Kagome, owing to its low dimensionality, are strong enough to destabilise the ferromagnetic order. The resultant non-magnetic metal is best described by the time reversal symmetric hopping Hamiltonian in eqn.~(\ref{oneorbHam}) along with the short range interactions in eqn.~(\ref{eq:H_int}).
Taking cue from the first principles, which predicts a large density of states at and close to the Fermi energy owing to  the presence of several nearly flat bands (see Fig.~3(c) of Ref.~\onlinecite{Baidya_PRL_2020}), we propose to work with a set of hopping parameters provided in \Fig{fig:lambda_bands} which captures this band phenomenology adequately. In the rest of this section, we shall discuss the effect of the short range interactions over these tight binding bands at appropriate fillings, while highlighting some of the interesting instabilities of the non-magnetic phase.

In the limit of large onsite repulsion and at electron densities of less than one per site, we can restrict to a space of electronic states which have no double occupancies. An appropriate parton paradigm in this limit splits the electron creation operator ($c_{i\sigma}^\dagger$) into a fermionic spinon creation operator ($f^\dagger_{i\sigma}$) and a bosonic hole annihilation operator ($b_i$)
such that 
\beq
c^\dagger_{i\sigma}=  f^\dagger_{i \sigma} b_i
\label{partontJ}
\eeq
with a local constraint 
\beq
\sum_\sigma f^\dagger_{i \sigma} f_{i \sigma} + b^\dagger_i b_i = 1\ \ \ \forall \ i.
\label{eq:p_constraint}
\eeq
Clearly, \eqn{partontJ} leads to a $U(1)$ gauge redundancy with the spinons and the holons carrying a unit charge each with respect to the corresponding gauge field \cite{2006_RMP_Lee}. This $U(1)$ gauge structure is in addition to the one associated with electromagnetism; the source of the interaction terms in eqn.~(\ref{eq:H_int}). To avoid any possible confusion we refer to the gauge structure implied by \eqn{partontJ} as the internal gauge structure. Furthermore, it is easy to convince oneself that within this parton construction the number of spinons at any given site must be identified with the number of electrons at that site, while the number of holons gives the probability of the site being empty.

\subsection{$t$-$J$ model}
A natural model to work with is an appropriate $t$-$J$ model defined in the no-double-occupancy sector of the electronic Hilbert space. In the presence of nearest neighbor interactions ($V_{ij}$) which are sub-dominant to the onsite interaction ($U$), this model Hamiltonian takes the following form
\begin{multline}
H_{tJ} =-\sum_{\stackrel{ij}{\sigma, \sigma'}} t^{\sigma \sigma '}_{ij} c_{i\sigma}^\dagger c_{j\sigma'}-\mu\sum_{i} n_i \\
+ \sum_{\langle ij\rangle} \left[ J_{ij} \bS_i \cdot \bS_j + \left( V_{ij} - \frac{J_{ij}}{4} \right) n_i n_j \right]
\label{tJmodel}
\end{multline}
where $S_{i}^{(\tau)} = \frac{1}{2} \sum_{ss'} c^\dagger_{i s}\sigma^{(\tau)}_{s s'} c_{i s'}$ with $\sigma^{(\tau)}$ being the Pauli matrices and $n_i = \sum_\sigma c_{i\sigma}^\dagger c_{i\sigma}^{}$ are, respectively, the electronic spin and number operators at site $i$. $t_{ij}^{\sigma \sigma'}$ are the same as in \eqn{oneorbHam} with values provided in \Fig{fig:lambda_bands} and $\mu$ is the chemical potential. $J_{ij}$ is the strength of the nearest neighbor anti-ferromagnetic exchange interaction that one obtains in the usual derivation of the $t$-$J$ model (see, for example, \cite{Chao_JPCSSP_1977,Hirsch_PRL_85}).
Note that in eqn.~\ref{tJmodel} we are not including the exchange processes arising from next nearest neighbour hopping or from the effective spin-orbit coupling as their contribution ($\sim g^2/U$ where $g$ is the amplitude of the hopping process) is much smaller compared to that from the nearest neighbor hopping processes. 

For the purpose of highlighting qualitatively non-trivial physics we find it sufficient to set $J_{ij} = J$ and $V_{ij}=V$ for all the nearest neighbor site indices $i$ and $j$. With these simplifications $H_{tJ}$ can be rewritten in a suggestive form as
\begin{multline}
    H_{tJ} = \sum_{ij} t^{\sigma \sigma '}_{ij} c_{i\sigma}^\dagger c_{j\sigma'}-\tilde{\mu}\sum_{i} n_i \\
    -J_\Delta \sum_{\langle ij \rangle} B_{ij}^\dagger B_{ij}^{} - J_\kappa \sum_{\langle ij \rangle}  K_{ij}^\dagger K_{ij}^{}
    \label{eq:tJ_rewrite}
\end{multline}
where $\tilde{\mu} = \mu - V$ is the renormalized chemical potential, $J_\Delta = J-2V$ and $J_\kappa = 2V$. $B_{ij}^\dagger = (c_{i\uparrow}^\dagger c_{j\downarrow}^\dagger - c_{i\downarrow}^\dagger c_{j\uparrow}^\dagger)/\sqrt{2}$ is the creation operator for a two electron singlet with one sitting on the $i$-th site and another on the $j$-th site, while $K_{ij}^\dagger = (c_{i\uparrow}^\dagger c_{j\uparrow} + c_{i\downarrow}^\dagger c_{j\downarrow})/\sqrt{2}$ is an operator which facilitates the hopping of an electron from site $j$ to site $i$. Clearly, if $J_\Delta > 0$, that is $J>2V$, a superconducting ground state is plausible at any electron filling of less than one electron per site. Conversely, if $J_\Delta < 0$ one would obtain a correlated metal. For $J_\Delta \sim 0$ one expects a quantum critical phase dominated by long-wavelength fluctuations of the superconducting order parameter over a metallic background.

In the following sub-sections we describe a superconducting phase of the non-magnetic bi-layer Kagome system and its zero temperature transition to a metallic phase within the parton paradigm discussed above. It is easy to convince oneself that the $J_\Delta$ and $J_\kappa$ terms in \eqn{eq:tJ_rewrite} have the same matrix elements in the electronic basis and in the spinonic basis even while leaving out the holon operators\cite{2006_RMP_Lee}. Therefore, for the purpose of streamlining the analysis that follows we shall write them completely in terms of the spinon operators. Moreover, in our analysis we shall relax the local constraints eqn.~(\ref{eq:p_constraint}) to a global constraint with the understanding that the physical wavefunctions can be obtained by projecting our wavefunctions to the space of physical states\cite{1995_PRB_Ichinose,2001_PRB_Ichinose}.

\begin{figure}
    \centering
\includegraphics[width=0.8\columnwidth]{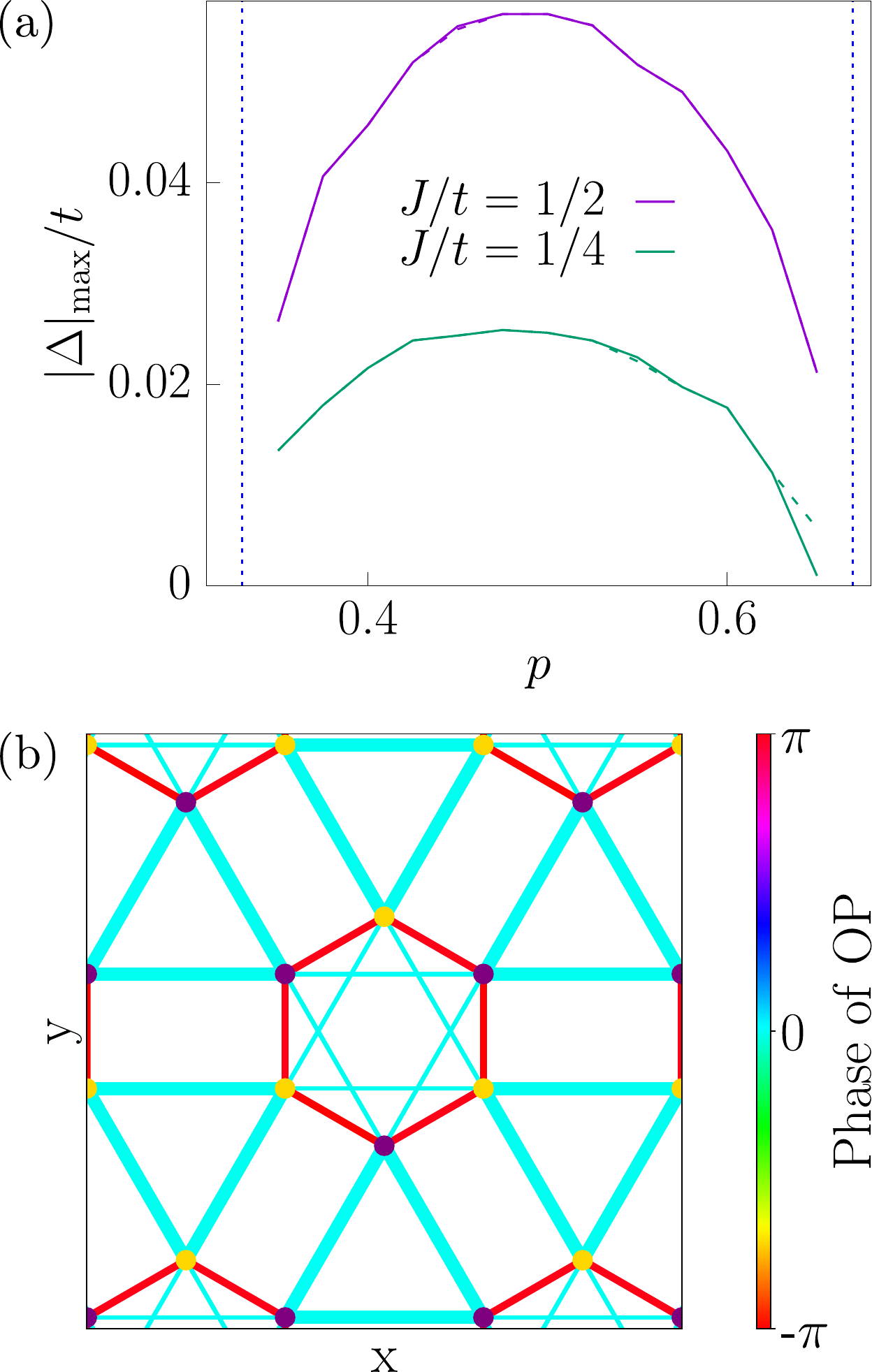}
    \caption{{\it SC within $t$-$J$ model:} (a) Behavior of $|\Delta|_{\text{max}}/t$ as a function of $p$ for $J/t = 1/2$ and $1/4$, obtained by solving t-J model in V=0 limit. Solid (dashed) curves correspond results obtained with a momentum space mesh of $151 \times 151$ $(101 \times 101)$ and the blue dotted lines mark the band edges at $p=1/3$ and $2/3$. (b) Corresponding real space variation of the superconducting order parameter at $J/t=1/4$ and $p=0.5$. The width of the bonds represent the magnitude of $\Delta$ (maximum $\sim 0.025t$ and and minimum $\sim 0.007t$) while the phase has been color coded.}
    \label{fig:tj}
\end{figure}

\subsubsection{High $T_c$ superconductivity}
For a generic electron density $(1-p)$ of less than one electron per site, the holons can safely be assumed to condense into a superfluid state with $\langle b_i \rangle \sim \sqrt{p}$. This results in a renormalization of the hopping amplitudes for the electronic quasi-particles, $t_{ij}^{\sigma \sigma'} \rightarrow p t_{ij}^{\sigma \sigma'}$ and thus captures the correct behaviour in the two limiting cases of $p\rightarrow0$ where one expects a Mott insulating ground state, and $p\rightarrow 1$ where one expects a Fermi liquid ground state. The condensation of holons further gaps out the internal gauge fields via the Anderson-Higgs mechanism and, by appropriately redefining the spinonic operators, allows us to work with electron-like operators\cite{1983_JPCSSP_Read}. In this background of condensed holons, we then obtain a mean field Hamiltonian $H_{tJ}^{MF}$ via the decomposition of the quartic terms in eqn.~(\ref{eq:tJ_rewrite}) in the pairing ($\{\Delta_{ij}\}$) and the kinetic channels ($\{\chi_{ij}\}$)
\begin{eqnarray}
    J_\Delta B^\dagger_{ij} B_{ij} & \simeq & \Delta_{ij}^* B_{ij} + B^\dagger_{ij} \Delta_{ij} - \frac{|\Delta_{ij} |^2}{J_\Delta} \label{eq:MF_decom_D}\\
    J_\kappa K^\dagger_{ij} K_{ij} & \simeq & \chi_{ij}^* K_{ij} + K^\dagger_{ij} \chi_{ij} - \frac{|\chi_{ij} |^2}{J_\kappa}. \label{eq:MF_decom_chi}
\end{eqnarray}
Finally, to solve $H_{tJ}^{MF}$ self-consistently we impose
\begin{eqnarray}
    \Delta_{ij} & = & J_\Delta \langle B_{ij} \rangle_{MF} \label{eq:sc_D}\\
    \chi_{ij} & = & J_\kappa \langle K_{ij} \rangle_{MF} \label{eq:sc_chi}
\end{eqnarray}
where $\langle \ \rangle_{MF}$ denotes that the expectation value is being computed in the ground state of $H_{tJ}^{MF}$. We also simultaneously adjust the chemical potential to accommodate $p$ electrons per site on an average.

Let us first discuss the superconducting phase that we obtain in the limit of vanishing nearest neighbor repulsion $V$ at $T=0K$. In this limit \eqn{eq:MF_decom_chi} and \eqn{eq:sc_chi} can be dropped, while $1/3<p<2/3$ is such that the non-interacting Fermi surface lies in the flat band (see \Fig{fig:lambda_bands}). The eighteen superconducting order parameters $\{\Delta_{ij}\}$ associated with a unit cell live on the bonds of the lattice (see \Fig{fig:tj}(c)) and can be divided into three disjoint sets on the basis of their transformation properties under the lattice symmetries. The bonds in any one of these sets mix under the lattice symmetries and thus, host superconducting order parameters of the same magnitude. The maximum magnitude of the pairing order parameter among these three sets, which should eventually decide the superconducting transition temperature, is defined to be $|\Delta|_{\text{max}}$ ($|\chi|_{\text{max}}$ is defined analogously).

$|\Delta|_{\text{max}}/t$ as a functions of $p$ has a dome shape with a vanishing trend close to the band edges $p=1/3$ and $2/3$ where the electronic density of states vanish. We show this in \Fig{fig:tj}(a) for $J/t = 1/2$ and $1/4$. In \Fig{fig:tj}(b) we plot $\{\Delta_{ij}\}$ over the lattice, with the thickness of the bond between sites $i$ and $j$ being proportional to the magnitude of $\Delta_{ij}$ and its phase being coded in the color of the bond. Clearly, the intra-layer and inter-layer pairing order parameters are out of phase while being invariant under lattice rotations, suggesting an $s_{\pm}$-wave symmetry of the pair wavefunction. In terms of absolute energy scales for $t\sim 0.1 eV$, $J\sim 0.05eV$ and $p=0.5$ we find a $|\Delta|_{max}$ of the order $5-6\%$ of $t$ which corresponds to a maximum $T_c \sim \langle b_i^\dagger b_{j}^\dagger \rangle |\Delta|_{max} \sim p |\Delta|_{max}$ of about $30-35 K$, which is rather high given that we are significantly away from a filling of one electron per site.

\subsubsection{Correlated metal}

The superconducting state that we obtained above seems quite robust and can in principle be the dominant instability of the nonmagnetic metal in the bi-layer Kagome system. However, if this superconducting phase could be suppressed some even more exotic phases might become realizable. Extensions of the Lieb-Schultz-Mattis theorem \cite{hastings2004lieb} suggest that at a filling of 3 electrons per unit cell ($p = 0.5$) if the electron-electron interactions are unable to stabilize an ordered phase with a broken symmetry then a quantum spin liquid ground state is possible. At this filling the band at the Fermi energy in Fig~\ref{fig:lambda_bands} is half filled and the corresponding Wannier orbitals would form a triangular lattice. On incorporating the effect of microscopic interactions appropriately one expects to obtain an effective tight binding model with one orbital per site and some short ranged interactions on the triangular lattice. The triangular lattice, like the Kagome lattice, is prone to frustrations and hence can be expected to host one of the exotic quantum spin liquid phases (see, for example, \cite{2020_PRX_Szasz}). Thus, it is pertinent to discuss the possible mechanisms of suppressing the superconducting order.

A mechanism for this suppression is provided by the fluctuations of the collective excitations about the superconducting state \cite{2020_PRL_Mallik} which may leave us with a correlated nonmagnetic metal. Increasing the nearest neighbor interaction strength $V$ offers yet another natural pathway for the suppression of the superconducting order in the bi-layer Kagome systems. Fig.~\ref{fig:SC_to_metal} shows the results of our mean-field calculations for $H_{tJ}$ in eqn~\ref{eq:tJ_rewrite} with non-zero values of $V$ at $J/t = 0.25$ and $p=0.4$. We find that as $V$ is increased towards $J/2$ the superconducting order parameter decreases monotonically and eventually vanishes via a first order quantum phase transition when $V_c \lesssim J/2$ to become a correlated metal as $V/t$ is increased further. A more detailed study uncovering the interesting instabilities, including ordering by symmetry breaking, of this correlated metal remain an interesting direction for future work.

\begin{figure}
    \centering
\includegraphics[width=1.0\columnwidth]{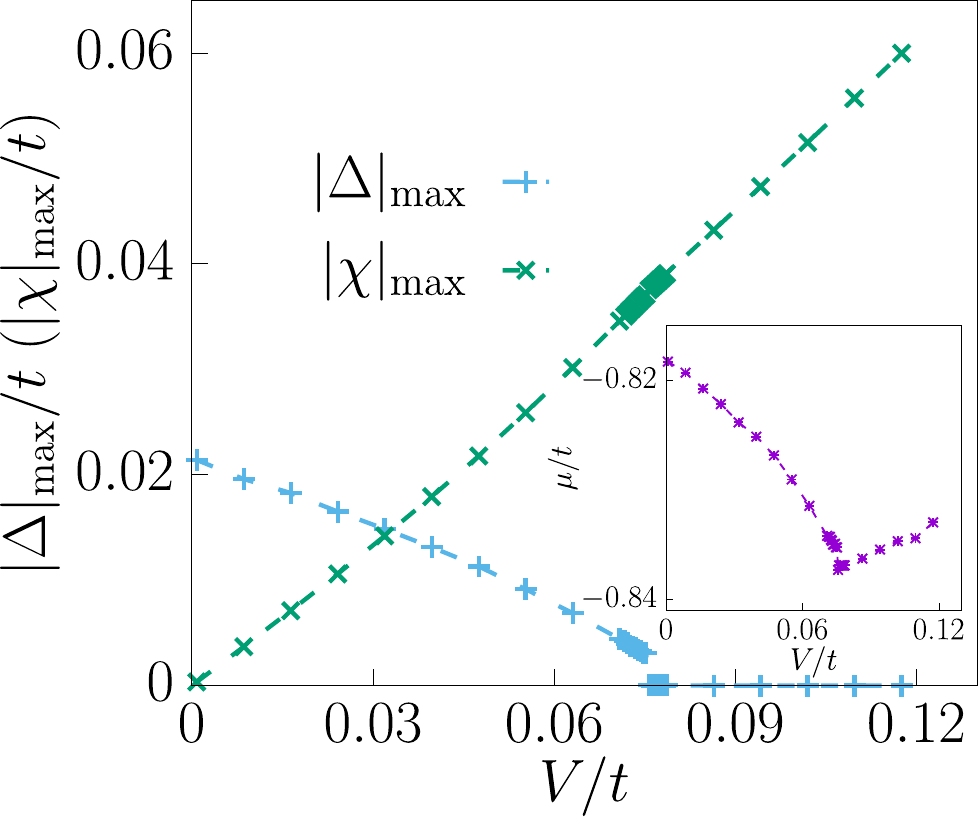}
    \caption{{\it SC to metal transition as a function of $V/t$}: Behavior of $|\Delta|_{\text{max}}/t$ and $|\chi|_{\text{max}}/t$ as a function of the nearest neighbor repulsion $V/t$ at $J/t=0.25$ and $p=0.4$. The inset shows the chemical potential as $V/t$ is tunned across the superconductor to metal transition at $V_c \lesssim J/2$. The momentum space mesh grid used for this calcuation was 151$\times$151.}
    \label{fig:SC_to_metal}
\end{figure}

\section{Summary and Discussion}
\label{compenergy}

Kagome intermetallics are an interesting set of materials which offers an ideal playground for studying the interplay of topology
and strong correlation effect. Motivated by the rich physics offered by
M$_3$Sn$_2$ class of these compounds, and possible experimental synthesis of
bi-layers of these materials, in this work, we investigate the correlation
driven instabilities in the Kagome bi-layer systems. Employing appropriate
numerical and analytical tools we solve the material-inspired model Hamiltonian.
We consider the two starting states, a) a Chern metal with spin-polarized bands
and b) a nonmagnetic metallic phase with magnetism being destroyed due to enhanced
fluctuations in two dimensions. We discover correlation-driven instabilities towards exotic phases, like fractional chern insulator, superconductivity and correlated 
metal. 

Turning on the correlation effect on Chern metallic phase, 
our analysis unravels possibility of two different FCI states. In the weak
inter-layer hybridization limit with the system hosting a flat band of Chern number $C=1$ near the Fermi energy, at one-third filling appears to stabilize 
a $1/3$ FCI state, while in the limit of strong interlayer hybridization where the system hosts a flat band with $C=2$ at the Fermi energy a $1/5$ FCI state may get stabilized.
The gap to excitations in the $1/3$ and $1/5$ FCI states are estimated to be $\sim0.02V$ and $0.002V$, respectively, (see \Fig{fig:1by3state} and \Fig{fig:1by5}), where $V$ is the inter-site interaction strength. Assuming a $V \sim 0.5eV$, this would correspond to the gap scaling of $\Delta^{\text{FCI}}_{1/3} \sim 10meV$ for the $1/3$ FCI and $\Delta^{\text{FCI}}_{1/5}\sim 1meV$ for the $1/5$ FCI, which are comparable to the width of the flat band in the bi-layer Kagome compound as predicted from first principles calculations\cite{Baidya_PRL_2020}. This suggests that to  achieve the observed FCI states in the real materials,
further band engineering may be needed to flatten out the bands at the Fermi energy. Also the separation of flat bands need to be ensured, demanding experimenting with choice of substrate, gating and strain engineering.

Consideration of strong repulsive interaction in the nonmagnetic situation, with restored time-reversal symmetry, leads to
distinctly different scenario. Here, using an appropriate $t$-$J$ model, we show that the system can realize a superconducting phase akin to high-temperature superconducting phases. Interestingly this superconducting phase undergoes a transition to a correlated metal as the inter-site repulsive interaction is increased. This further opens up the possibility of realizing a quantum spin liquid phase when the band at the Fermi energy is half filled. The precise nature of the superconductor to the correlated metal transition, beyond mean-field approximation, and similarly the character and the instabilities of the correlated metal are interesting directions for future work.

It is important to note that in our work we have considered
only the effect of repulsive interaction. As shown in Ref.~\onlinecite{Baidya_PRL_2020}, an attractive interaction in Chern metallic phase can lead to a topological superconducting phase. This might have phononic origins as has recently been seen in some transport experiments in antimony based compounds \cite{LuoarXiv2021}. Out of the several possible instabilities,
which one will dominate will depend upon the material specific details of
the band structure, the strength and nature of short-ranged electron-electron correlation and electron-phonon interactions. Tuning mechanisms available due to the layered and
cleavable nature of the systems, though, make the situation promising, opening
up the door for experimental exploration. Using the substrate effect, straining
and gating as handles we envisage a rich phase diagram to emerge in this
rich class of compounds.


\acknowledgements

We acknowledge Subhro Bhattacharjee for extensive discussions related to this work and collaborations on related projects and ideas. We acknowledge use
of open-source QuSpin[77, 78] for exact diagonalisation
calculations. AA acknowledges support from IITK Initiation Grant (IITK/PHY/2022010). We acknowledge the hospitality of ICTS, Bangalore where a significant part of this work was done. During this time AVM was also supported by a postdoctoral fellowship of ICTS, Bengaluru. T.S-D acknowledges J.C.Bose National Fellowship (grant no. JCB/2020/000004) for funding.

\appendix

\begin{table}
\centering
\begin{tabular}{|c||c|c|c|c|c|c|}
	\hline
	\diagbox[width=4em]{dof's}{Symm.\\Ele.} & $\mathrm{E}$ & $\mathrm{C}_3$ & $\mathrm{C}'_2$ & $\mathrm{i}$ & $\mathrm{S}_6$ & $\sigma_d$\\
	\hhline{|=||=|=|=|=|=|=|}
	1 & 1 & 2 & $1'$ & $1'$ & $3'$ & 1\\
	\hline
	2 & 2 & 3 & $3'$ & $2'$ & $1'$ & 3\\
	\hline
	3 & 3 & 1 & $2'$ & $3'$ & $2'$ & 2\\
	\hline
	$1'$ & $1'$ & $2'$ & 1 & 1 & 3 & $1'$\\
	\hline
	$2'$ & $2'$ & $3'$ & 3 & 2 & 1 & $3'$\\
	\hline
	$3'$ & $3'$ & $1'$ & 2 & 3 & 2 & $2'$\\
	\hline
	$\Gamma$ & $\Gamma$ & $\Gamma$ & $\Gamma$ & $\Gamma$ & $\Gamma$ & $\Gamma$\\
	\hline
	$K$ & $K$ & $K$ & $K$ & $K'$ & $K'$ & $K'$\\
	\hline
	$K'$ & $K'$ & $K'$ & $K'$ & $K$ & $K$ & $K$\\
	\hline
	$M$ & $M$ & $M''$ & $M$ & $M$ & $M'$ & $M$\\
	\hline
	$M'$ & $M'$ & $M$ & $M''$ & $M'$ & $M''$ & $M''$\\
	\hline
	$M''$ & $M''$ & $M'$ & $M'$ & $M''$ & $M$ & $M'$\\
	\hline
\end{tabular}
	\caption{{\it Symmetry analysis:} This table shows how different real space (1, 2, 3, $1'$, $2'$ and $3'$ are the sites in a unit cell; see Fig.~\ref{fig:figure1}) and reciprocal space ($\Gamma$, $K$, $K'$, $M$, $M'$ and $M''$ are high symmetry points in the hexagonal Brillouin Zone; see inset of Fig.~\ref{fig:lambda_bands}) degrees of freedom (dof's) transform under a representative element of each of the conjugacy classes of the $\mathrm{D}_{3\mathrm{d}}$ point group (see, for example, Ref.~\onlinecite{jacobs_GT}).} 
	\label{tab:reducbl_d3d}
\end{table}

\section{The low energy tight-binding models}
\label{appen_tb}

\begin{figure*}
\includegraphics[height=0.8\columnwidth]{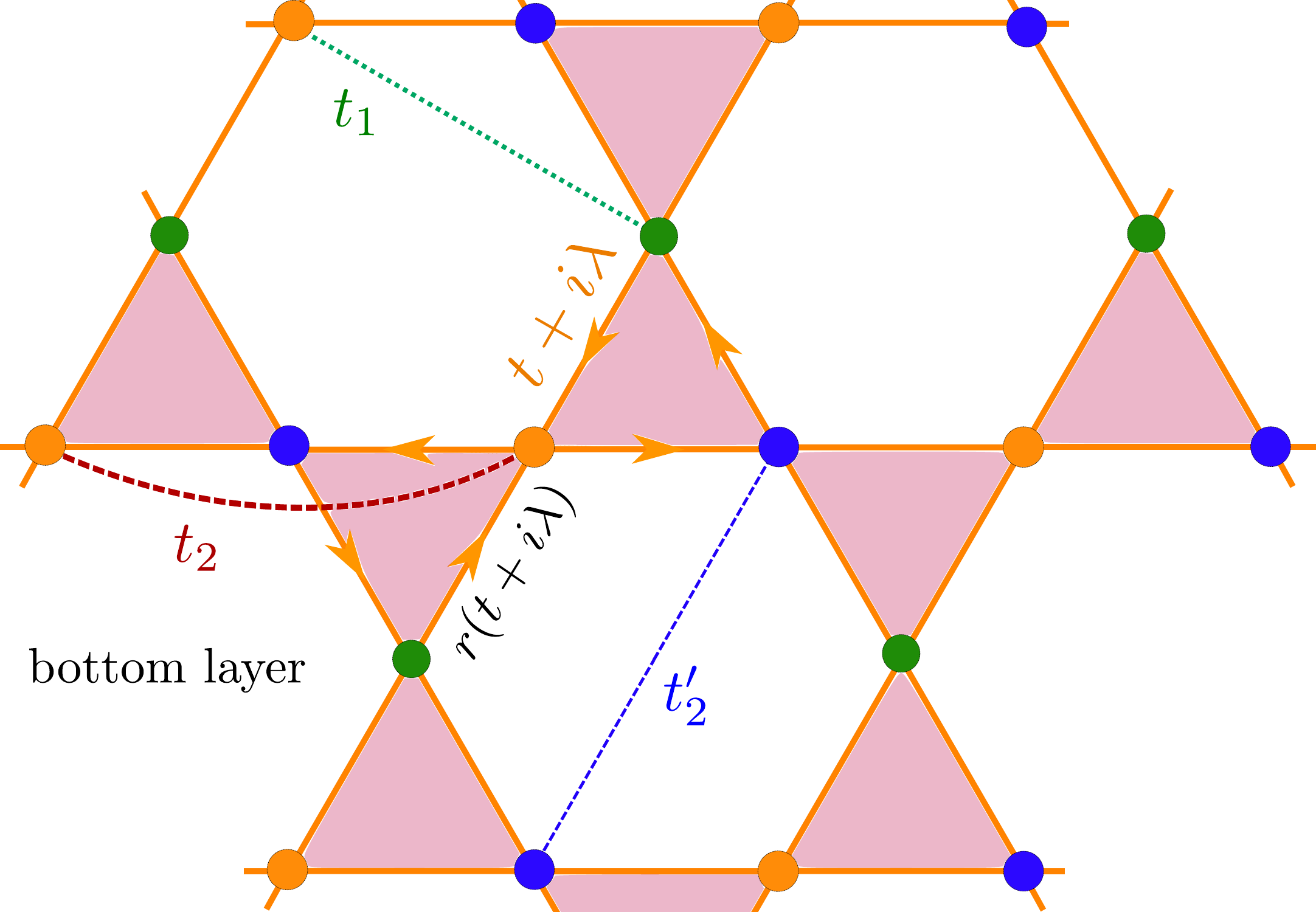}
\includegraphics[height=0.8\columnwidth]{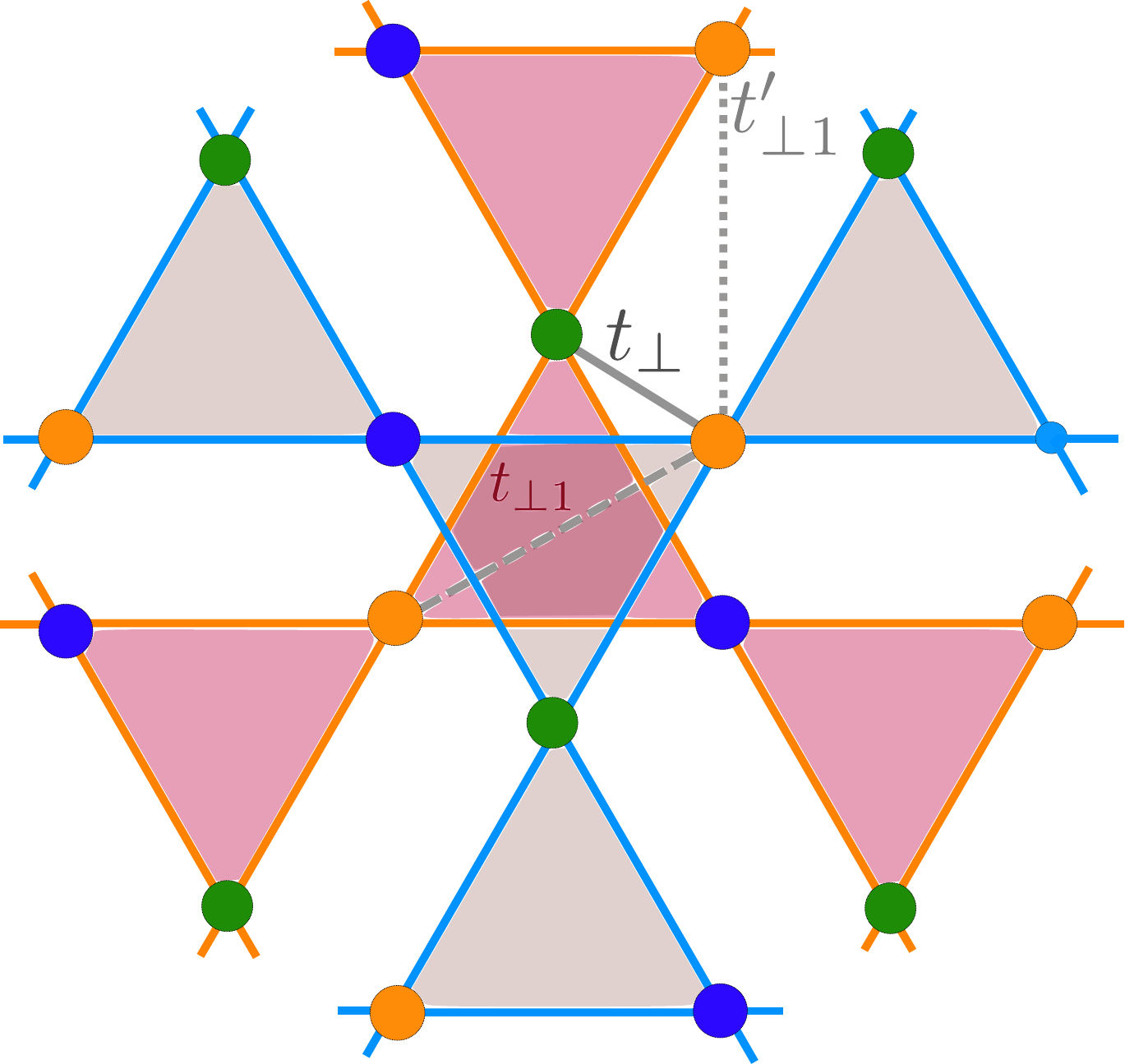}
\caption{{\it  Hopping processes:} The left (right) panel shows a portion of the lower Kagome layer (bi-layer Kagome system) and also depicts all the intra-layer (inter-layer) hopping amplitudes for up-spin electrons. Rotations of these schematic diagrams by $120^\circ$ about an axis perpendicular to the plane and passing through the center of the central unit cell, generates all the hopping processes associated with this unit cell. Furthermore, the hopping structure in the upper Kagome layer can be obtained by rotating the left schematic by $180^\circ$ about an axis perpendicular to the plane and passing through the centre of an up triangle. Finally, the hopping amplitudes for the down-spin electrons can be obtained by switching the sign of $\lambda$. Note that all the Kagome sites are equivalent and different colors have only been used to highlight the tripartite structure.}
\label{fig:hopping_params}
\end{figure*}

The point group symmetry of the bi-layer kagome lattice (see \Fig{fig:figure1}) is $\mathrm{D}_{3\mathrm{d}}$
(see Table~\ref{tab:reducbl_d3d}) and the symmetry consistent short ranged hopping processes, which we consider are shown in \Fig{fig:hopping_params}. Given that the spin dependent hopping in Eq. \ref{eq_soc} does not mix the spin-flavours we can still use $S_z=\uparrow,\downarrow$ as a good quantum number. We now present the symmetry analysis of the bands for both $\lambda=0$ and $\lambda \neq 0$ separately.

When $\lambda=0$, the little group at the $\Gamma$ point of Brillouin zone is $\mathrm{D}_{3\mathrm{d}}$.
From Table~\ref{tab:reducbl_d3d} one finds the reducible character vector of this group representation $\mathrm{R}_{\Gamma}$ to be $\left(6, 0, 0, 0, 0, 2\right)$ which implies $\mathrm{R}_\Gamma = \mathrm{A}_{1\mathrm{g}} \bigoplus \mathrm{E}_{\mathrm{g}} \bigoplus \mathrm{A}_{2\mathrm{u}} \bigoplus \mathrm{E}_{\mathrm{u}}$ implying two non-degenerate and two doubly degenerate states at the $\Gamma$ point. 
The little group of the $K$ point includes the transformations $\mathrm{C}_3$ and $\mathrm{C}_2'$
which generate the discrete group $\mathrm{D}_{3}$
(see Ref.~\onlinecite{jacobs_GT}). The corresponding transformation matrices can be read off from Table~\ref{tab:reducbl_d3d} and they together form a representation $\mathrm{R}_K$ of $\mathrm{D}_{3}$. The character vector for $\mathrm{R}_K$ turns out to be $(6,0,0)$ implying that $\mathrm{R}_K = \mathrm{A}_1 \bigoplus \mathrm{A}_2 \bigoplus 2\mathrm{E}$. Therefore, the degeneracies in the spectrum at the $\mathrm{K}$ point is exactly the same as that at the $\Gamma$ point discussed above. The little group of the $M$ point is the Abelian group $C_i$ and its reducible representation given in Table~\ref{tab:reducbl_d3d} $R_M$, has $(6,0)$ as its character vector. Consequently, the spectrum at the $M$ point is non-degenerate with $\mathrm{R}_M = 3\mathrm{A}_\mathrm{g}\bigoplus 3\mathrm{A}_\mathrm{u}$.

For $\lambda \neq 0$, the point group symmetry of each spin sector changes from $\mathrm{D}_{3\mathrm{d}}$ to $S_6$ which only has one dimensional irreducible representations 
(see Ref.~\onlinecite{jacobs_GT}). The character vector of this representation at the $\Gamma$ point is $(6,0,0,0,0,0)$ (see Table~\ref{tab:reducbl_d3d}) which means a completely non-degenerate spectrum. Similarly, the little group at the $K$ point is $\mathrm{C}_3$ (see Ref.~\cite{jacobs_GT})
which admits only one dimensional irreps, implying a non-degenerate spectrum, with each irrep appearing twice. 

The change in the dimension of the irreps in the presence of $\lambda$ as compared to those in its absence is evident from the dispersion at the $\Gamma$ and the $K$ points as sown in Fig.~\ref{fig:lambda_bands}. We note that the bands at every momentum point has an additional two-fold degeneracy, which is guaranteed by a combination of time-reversal and inversion.

The inversion symmetry, however, may generally be absent in an experimental realization because of the presence of a substrate or an applied gate potential. When $\lambda=0$, the residual point group symmetry of a given $S_z$ sector is $\mathrm{C}_{3\mathrm{v}}$ (see, for example, Ref.~\cite{jacobs_GT}). The reducible representation (see Table~\ref{tab:reducbl_d3d}) at the $\Gamma$, has the character vector $(6,0,2)$, implying the following decomposition in terms of the irreducible representations of $\mathrm{C}_{3\mathrm{v}}$, $2\mathrm{A}_1 \bigoplus 2\mathrm{E}$. The little groups at the  $K$ and the $M$ points are, respectively, $\mathrm{C}_3$ and $\mathrm{C}_1$, which are both Abelian. When $\lambda\neq0$, the point group symmetry of the system reduces to $\mathrm{C}_3$ implying generically non-degenerate bands at all momenta in the Brillouin zone. When both the $S_z$ sectors are considered such that time-reversal $\mathrm{T}$ with $\mathrm{T}^2=-\mathbb{1}$ is also present, the band structure exhibits Kramer's degeneracy.

\section{Mean field ferromagnetism}
\label{sec:mft_fm}
A self-consistent mean field analysis of the onsite repulsive term (see \eqn{eq:U_m}) for the free dispersion given in \Fig{fig:lambda_bands} results in fully spin-polarized band at low energies for $U/t \gtrsim 1$ (see \Fig{fig:U_vs_m}).
\begin{figure}
    \centering
    \includegraphics[width=0.8\columnwidth]{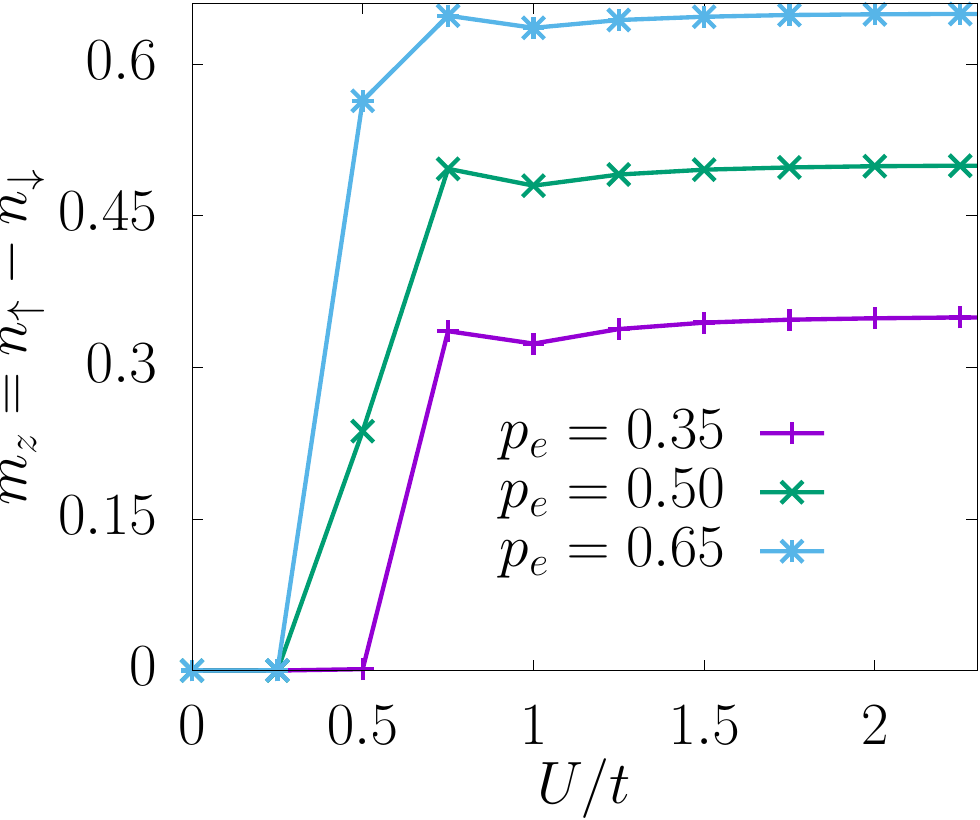}
    \caption{{\it Mean field ferromagnetism:} For the bands presented in Fig.~\ref{fig:lambda_bands}, we plot the average magnetization per site ($m_z = 1/N_s \sum_i  (n_{i\uparrow} - n_{i\downarrow}) $) against the strength of onsite Hubbard interaction $U/t$ for different values of electron filling per site $p_e$ ($V_{ij}=0$, see \eqn{eq:H_int}). Fully spin polarized state is realized for $U/t \gtrsim 1.0$.}
    \label{fig:U_vs_m}
\end{figure}

\section{Absence of superconductivity with repulsive interactions for Chern metal}
\label{SCspinpola}

\begin{figure}
\centering
    \includegraphics[width=0.8\linewidth]{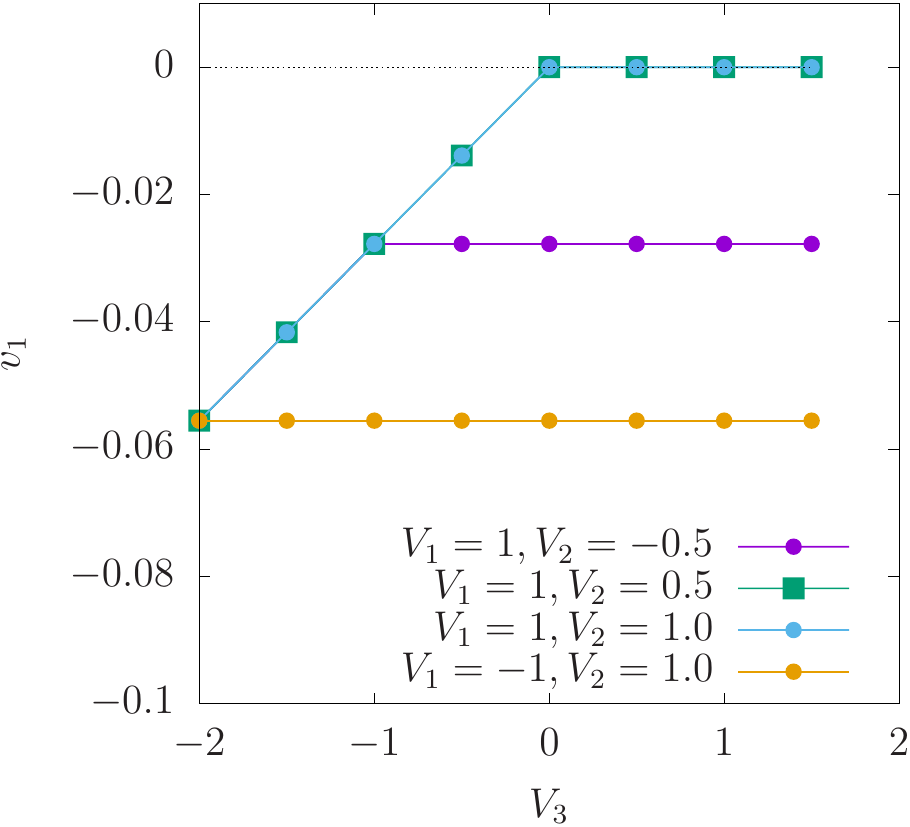}
	\caption{{\it Absence of SC in Chern metal:} Variation of $v_1\ (=\{v_n\}_{min})$ for some representative values of $V_1$, $V_2$ and $V_3$, showing that when $V_1,V_2,V_3>0$ the system does not host any $v_n<0$.}
	\label{fig:repulsiveV}
\end{figure}

The quartic term in a many body Hamiltonian which describes the scattering of two fermion pairs, each with a zero center-of-mass momentum, is relevant for the formation of Cooper pairs. The zero-momentum scattering term is given by
\begin{equation}
H_{p} = \sum_{\stackrel{k,k'}{s_1,s_2,s_3,s_4}} V_{k,s_1,s_2;k',s_3,s_4} c_{ks_1}^\dagger c_{-ks_2}^\dagger c_{-k's_4}c_{k's_3}
\label{eqn:Hampert}
\end{equation}
where $k$ and $k'$ are momentum indices, $s_{1(234)}$ denote both the atomic sites within the unit cell and also the spins. $V_{k,s_1,s_2;k',s_3,s_4}$, owing to fermionic statistics, satisfies   $V_{k,s_1,s_2;k',s_3,s_4} = -V_{-k,s_2,s_1;k',s_3,s_4} =  -V_{k,s_1,s_2;-k',s_4,s_3} = V_{-k,s_2,s_1;-k',s_4,s_3}$. 
Whether a zero-momentum pairing in any angular-momentum channel is favoured can be found by rewriting \eqn{eqn:Hampert} as 

\begin{eqnarray}
    H_p &=& \sum_{\stackrel{k,k'}{s_1,s_2,s_3,s_4}} c_{ks_1}^\dagger c_{-ks_2}^\dagger V_{k,s_1,s_2;k',s_3,s_4} c_{-k's_4}c_{k's_3} \nonumber \\
    &\equiv& \quad \quad \sum_n v_n \hat{\Delta}^\dagger_n \hat{\Delta}_n
    \label{quartterm}
\end{eqnarray}
where $v_n$ is $n$-th eigenvalue of matrix $V$.
The numbers $\{v_n\}$ denote the bound state energies of the $n^{th}$ zero momentum fermion pair where $\Delta_n$ is given by 
\begin{equation}
\label{eq:pair_wf}
\hat{\Delta}_n = \sum_{k',s_3,s_4} \mathrm{U}^\dagger_{n;k',s_3,s_4} c_{-k's_4}c_{k's_3}   
\end{equation}
such that $\mathrm{U}$ diagonalizes \eqn{quartterm}. Under a self-consistent Hartree fock, different pairing amplitudes $\Delta_n$ may take a finite expectation value. When the Fermi surface is nearly circular a superconducting instability in the $n^{th}$ channel is expected if $v_n <0$ \cite{2005_AIP_Sigrist}.

When such an analysis is performed for the ferromagnetic Chern metallic band, for three kinds of repulsive interactions: (i) $V_1$: intra-layer intra-unit cell nearest neighbor density-density interaction, (ii) $V_2$: intra-layer inter-unit cell nearest neighbor density-density interaction and (iii) $V_3$: inter-layer intra-unit cell nearest neighbor density-density interaction, we find no window in the parameter space for ($V_1,V_2,V_3>0$) when $v_n<0$ signalling a BCS state cannot be realized within a self-consistent mean field theory for purely repulsive microscopic interactions. Variation of $v_1\ (=\{v_n\}_{min})$ for some representative values of $V_1$, $V_2$ and $V_3$ is shown in \Fig{fig:repulsiveV}.

\begin{figure}
\centering
\includegraphics[width=0.8\linewidth]{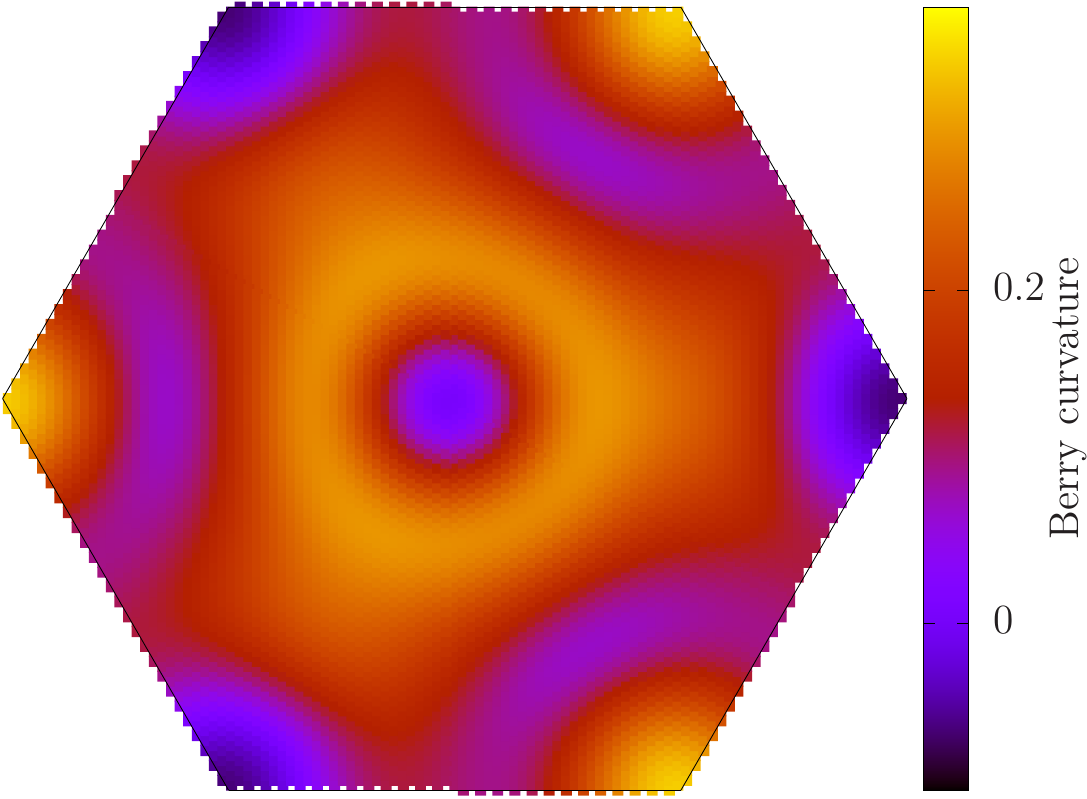}
	\caption{{\it Berry curvature:} Berry curvature distribution over the hexagonal Brillouin zone for the $C=1$ band at the Fermi energy in \Fig{fig:1byn_bands}(a)}
	\label{fig:berry_1by3}
\end{figure}
\begin{figure}
\centering
\includegraphics[width=0.8\linewidth]{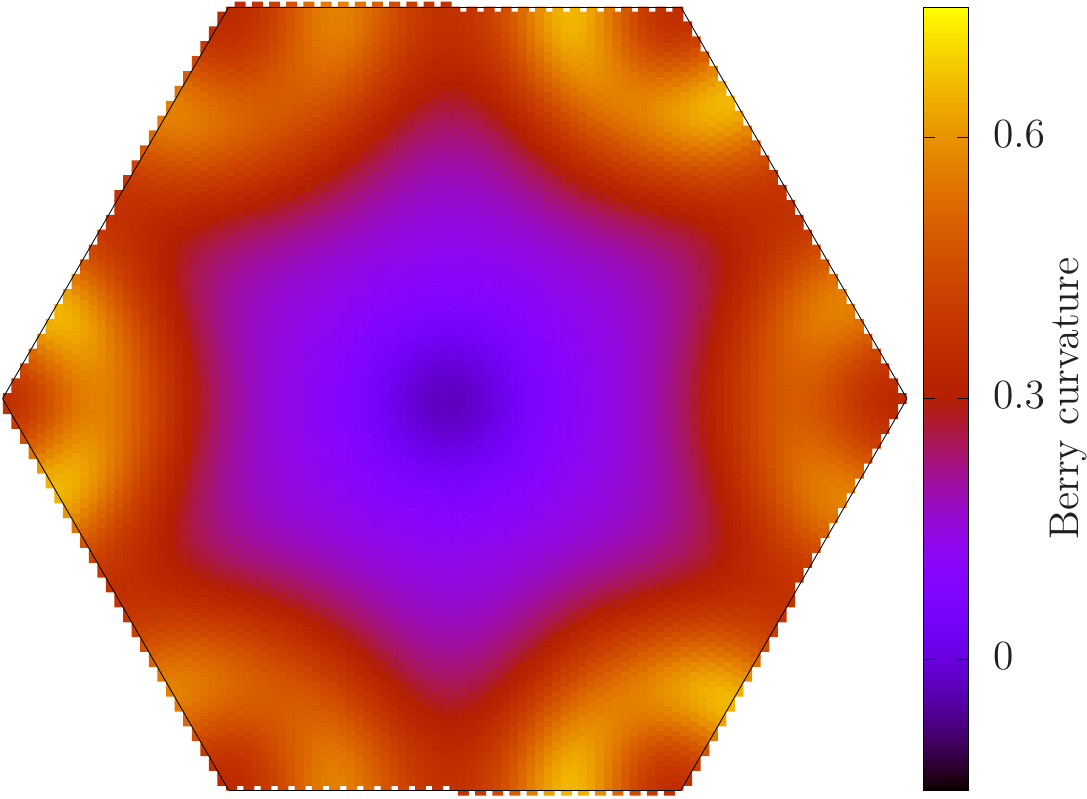}
	\caption{{\it Berry curvature:} Berry curvature distribution over the hexagonal Brillouin zone for the $C=2$ band at the Fermi energy in \Fig{fig:1byn_bands}(b)}
	\label{fig:berry_1by5}
\end{figure}

\begin{figure}
\centering
    \includegraphics[width=0.8\linewidth]{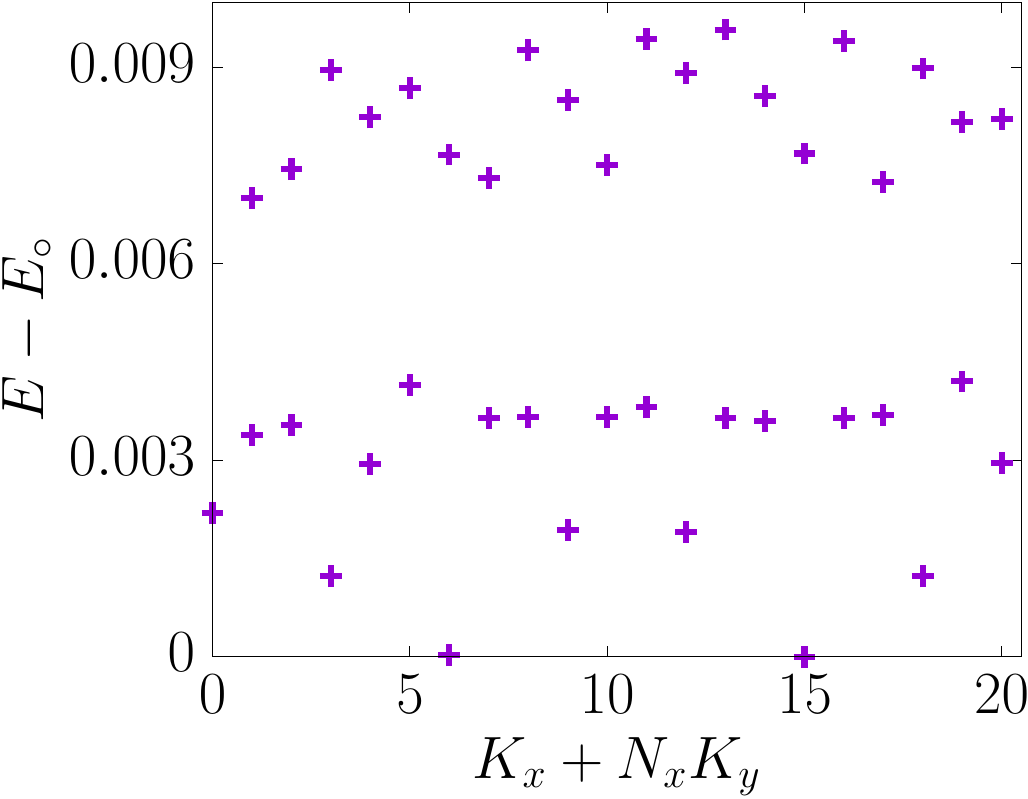}
	\caption{{\it 1/5 FCI Quasihole spectrum:} The many-body spectrum for the flattened low energy band in \Fig{fig:1byn_bands}(b) with one quasi-hole about 1/5 FCI state; $3\times7$ system with 4 particles.}
	\label{fig:one_qh_1by5}
\end{figure}

\section{Fractional Chern Insulator}
\label{FCIextra}

Here we present additional results pertaining to the FCI states.

{\it FCI 1/3}: The Berry curvature over the hexagonal Brillouin zone for the $C=1$ band at the Fermi energy in \Fig{fig:1byn_bands}(a) is shown in \Fig{fig:berry_1by3}.

{\it FCI 1/5}: The Berry curvature over the hexagonal Brillouin zone for the $C=2$ band at the Fermi energy in \Fig{fig:1byn_bands}(b) is shown in \Fig{fig:berry_1by5}. In addition, the one quasi-hole spectrum is shown in \Fig{fig:one_qh_1by5} and number of states below the gap follows the expected counting arguments \cite{Regnault_PRX_2011}.

\bibliography{FCI}

\end{document}